# Approximation of the electronic terms of diatomic molecules by the Morse function. The role of anharmonicity. II. Simple terms

G.S. Denisov, R.E. Asfin*

*Department of Physics, Saint Petersburg State University, 7/9 Universitetskaya Nab., 199034 Saint Petersburg, Russian Federation*

*Corresponding author: R.Asfin@spbu.ru

**Abstract**

This article continues the series of works by the authors on the approximation of the electronic terms of diatomic molecules by the Morse formula, which is the simplest anharmonic approximation of the real term $U(r)$. Depending on the choice of parameters, the approximation has two alternative solutions $M1(r)$ and $M2(r)$, with different patterns of deviations from the real term and its vibrational structure. The difference $\delta(r) = U(r) - M(r)$ quantitatively shows the changes in the shape of the terms during approximation. We introduced an empirical anharmonicity function $-2\omega_e x(v)$, which characterizes the positions of vibrational levels in the potential well; it demonstrates the distortion of the vibrational structure of the term $U(r)$ during the approximation. Based on the data from literature, the functions $\delta(r)$ and $-2\omega_e x(v)$ were constructed for more than 20 molecules. Here we present a group of simple terms with minimal deviations from the Morse shape.

## 1. Introduction

Despite the rough approximation, the high popularity of the three-parameter Morse function is beyond doubt, mainly due to the possibility of analytically solving the Schrödinger equation with this potential with a good accuracy. Applications of the Morse approximation are rapidly growing in fundamental and applied physical chemistry, in particular, in the study of intermolecular interactions, adsorption, impurity centers in crystals, elementary reaction kinetics, etc. Therefore, the development of Morse's ideology and methods of its practical application are relevant.

In the current communication the main attention is focused on the development of the original results of the authors' recent work on the features of the Morse approximation [1–4]. They include the existence of two independent interpretations of the solution of the Schrödinger equation $M1(r)$ and $M2(r)$ for an alternative approximations of the potential function of the molecule $U(r)$, the introduction of the empirical anharmonicity function $-2\omega_e x(v)$ *(vide infra, Eq.(6))*, which characterizes the anharmonicity of the potential well in the vicinity of the vibrational quantum number $v$, the development of the approach for estimating the deviations of the models $M1(r)$ and $M2(r)$ from the real term $U(r)$. To estimate the shape of the contour of the models and the structure of their vibrational spectra, the approach involves the combined utilization of the Birge-Sponer plot and the difference $U(r)$ and model functions, $\delta(r) = U(r) - M(r)$. A primary empirical classification of electronic terms is proposed based on the shape of the anharmonicity function $\omega_e x(v)$ and deviations $\delta(r)$ of $M1(r)$ and $M2(r)$ from the true potential. The main results were briefly described in the preprint [4], here we describe the simple terms for the increased number of examples.

## 2. Approach

The Morse formula used to approximate the real electronic term of a diatomic molecule has the form [5]:

$$M(r) = D_e\left[1 - \exp\left(-a(r - r_e)\right)\right]^2 \qquad (1),$$

the eigenvalues (vibrational energies) of this potential are equal approximately (series is limited by two first terms):

$$E(v)/hc = G(v) = \omega_e(v + ½) - \omega_e x_e (v + ½)^2 \qquad v = 0,1,2 \ldots v_m \qquad (2),$$

where $\omega_e$, the harmonic frequency, $cm^{-1}$, and $\omega_e x_e$, anharmonicity, $cm^{-1}$, are parameters which can be calculated from two experimental vibrational frequencies $\omega(0-1)$ and $\omega(0-2)$. $v_m$ is the maximum vibrational quantum number, which defines the bond energy $D_e$. The coefficients in Eq. (2) depend on the parameters of Eq. (1):

$$\omega_e = a(2D_e/\mu)^{½} \qquad (3a), \qquad \omega_e x_e = a^2/2\mu \qquad (3b).$$

where $\mu$ represents the reduced mass of the diatomic molecule. Excluding "$a$", one obtains the expression for the bond energy

$$D_e = \omega_e^2/4\omega_e x_e \qquad (4).$$

It defines $D_e$ as the distance from the bottom of the potential Morse curve, Eq. (1), to the asymptote. This formula is exact only if Eq. (2) is strictly correct, and the positions of the vibrational levels are described by a single anharmonicity coefficient $\omega_e x_e$. In reality, when applying Eq. (4) to real terms, its inexactness always complicates the description of the system.

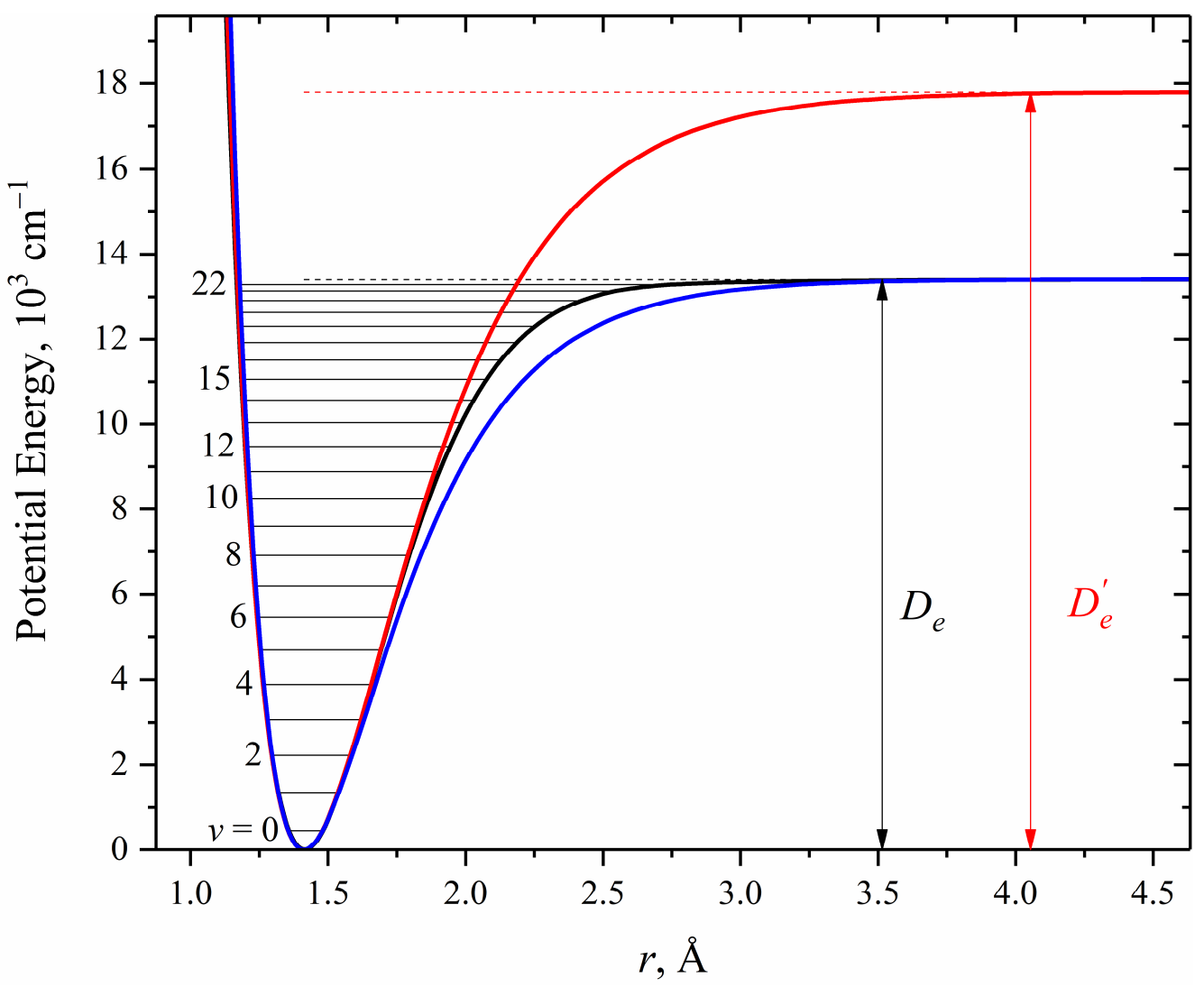

**Figure 1**. The term X $^1\Sigma_g^+$ of the $F_2$ molecule (black line) according to [9] and its Morse approximations $M1(r)$ (red) and $M2(r)$ (blue). $D_e$= 13408.49 $cm^{-1}$, $D_e'$ = 17780 $cm^{-1}$.

It was shown in [1] that when approximating the real term $U(r)$ by the Morse function, the exponent "$a$" in Eq. (1) can be expressed in terms of parameters $D_e$ and $\omega_e x_e$ in two ways, using either Eq. (3a) or Eq. (3b). The resulting solutions $M1(r)$ and $M2(r)$ are not identical [3], characteristic distortions are shown in Fig.1 for the ground-state electronic term of the $F_2$ molecule (black curve). Its approximation $M1(r)$ is shown by a red line and $M2(r)$ by a blue line. The $M1(r)$ term is located above the real term, part of it is in the region of the continuous spectrum, the $M2(r)$ term lies below, the deviation is not monotonous with maximum in the upper part of the potential curve. A detailed description of the results of this approximation is given below. The choice of the type of approximation is determined by the kind of the problem, – $M1(r)$ is often used to estimate the binding energy, $M2(r)$ allows for the use of the vibrational structure in the studies of laser pumping, thermodynamic characteristics of the ensemble, etc. An interesting example of the rarely used $M2(r)$ has recently appeared [6]. Such ambiguity is inherent in Morse's formula from the very beginning, it is explained by the impossibility of reaching its goal using one parameter - to introduce into the harmonic potential two independent parameters that define the elementary anharmonic oscillator – the asymptote $D_e$, and the anharmonicity $x_e$ (or $\omega_e x_e$).

Eq. (3a) defines $D_e$, but then one needs to fix $\omega_e x_e$, and this yields $M1(r)$; Eq. (3b) gives $\omega_e x_e$, but in this case, one needs to know $D_e$, and this gives $M2(r)$. For $M1(r)$, the $D_e'$ value (following [7] we will mark the calculated parameters of Morse models with primes) is obtained by a distant extrapolation based on the anharmonicity of the lowest vibrational levels, and for $M2(r)$, for a given $D_e$ and $\omega_e$, the dimensionless anharmonicity $x_e$ is calculated from Eq. (4) as $x_e' = \omega_e/4D_e$. The existence of two different approximations, distinguished by the choice of Eq. (3a) or Eq. (3b), was previously noted in the article [8].

It can be supposed that the main Morse's goal was to determine the bond energy, and the use of Eq. (4) was a necessary intermediate step. The possibility of an alternative solution for $M2(r)$ follows from Morse's construction, but was not explicitly mentioned by him.

The purpose of this work is to find out the appearance of the distortions of the initial potential function and its vibrational structure occurring when approximated by the Morse formula, to quantify the magnitude of the distortions by the $M1(r)$ and $M2(r)$ models using specific examples, and to try to find empirical possibilities for the primary systematization of the results obtained.

### 2.1. Quantitative characteristic of Morse approximations

Let us consider as an example a molecule for which the shape of the deviation of the electronic term from the Morse model has the simplest form. Fig. 1 shows the term of the X $^1\Sigma_g^+$ ground state of the $F_2$ molecule with vibrational levels constructed from the high-quality calculated data [9]. Based on the same data, approximations of this term by the Morse formula $M1(r)$ (red line) and $M2(r)$ (blue line) were constructed. The energy of the $M1(r)$ term is almost always significantly overestimated compared to $U(r)$, this was noted by Morse and subsequent authors ([5,10–12]). This means that the vibrational levels in Eq. (2) with given anharmonicity Eq. (3a) converge more slowly than in $U(r)$. The asymptote for $M1(r)$ is much higher than that for the real term, and due to distant extrapolation, the number of vibrational levels increases significantly. A smaller number of excess levels also appears in the $M2(r)$ function due to the increased anharmonicity coefficient $\omega_e x_e'$, so that the Morse approximation is always accompanied by an increase in the number of vibrational levels.

The extrapolated value of the energy $D_e'$ =17780 cm$^{-1}$ is by 33% higher than $D_e$=13408.49 cm$^{-1}$ [9]. Curve $M2(r)$ has an asymptote equal to the real value of $D_e$, and the anharmonicity $\omega_e x_e'$ must be greater than $\omega_e x_e$. The density of vibrational levels increases, as well as the width of the potential well, so the curve $U(r)$ runs between $M1(r)$ and $M2(r)$ without crossing. The distance between $U(r)$ and $M2(r)$ increases at first, passes through a maximum and then decreases to zero when $U(r)$ and $M2(r)$ reach the common asymptote $D_e$.

A clear and comprehensive characteristic of approximation of the term $U(r)$ is the difference – the deviation function $\delta(r) \equiv U(r) - M(r)$ for $M1(r)$ and $M2(r)$ [3]. Fig. 2 shows plots of these data for the $F_2$ molecule, built according to the results of [9]. Fig. 2 also provides information about the vibrational structure of three potential curves $U(r)$, $M1(r)$ and $M2(r)$, supplementing the data in Fig. 1. In the panel a) the dashes show the positions of the abscissa of the classical external turning points $r_2$ for the attraction branch in the coordinates of Fig. 1. In the panel b), the ordinates of these points are shown by dashes, so the presented data characterize quantitatively a system of vibrational levels of terms $M1(r)$ and $M2(r)$.

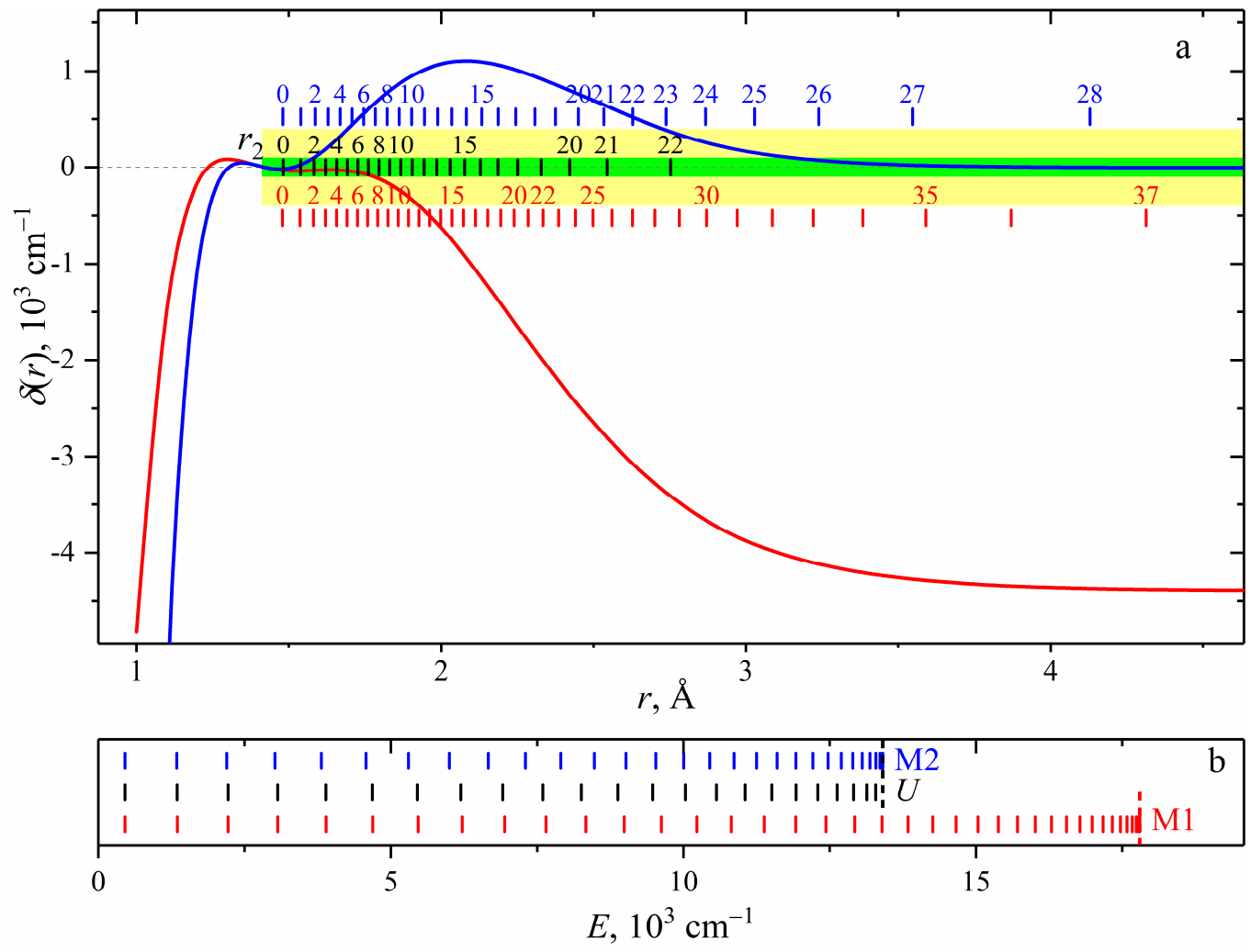

**Figure 2**. a. Deviations of $M1$ (red line) and $M2$ (blue) from the term X $^1\Sigma_g^+$ of the $F_2$ molecule according to [9]. Vertical dashes near the level of $\delta(r) = 0$, shown by the dashed line, are the values of the outer classical turning points $r_2$ of vibrational levels for potential curves $U(r)$ (black), $M1(r)$ (red), and $M2(r)$ (blue). The abscissa scale of $r_2$ coincides with the abscissa scale of the function $\delta(r)$. b. The energies of vibrational levels respect to the minima of the potential for $U$ (black), $M1$ (red), and $M2$ (blue). There are 16 fictitious levels for $M1(r)$ and 7 for $M2(r)$.

In the lower part of the potential well (~40% of its depth, up to $v = 7$ in Fig.1), the attraction branch of M1 with an excess of no more than 100 cm$^{-1}$, is close to $U(r)$. This range of precision better than 100 cm$^{-1}$ is colored green, the acceptable precision range 400 cm$^{-1}$ is colored yellow. For $M2(r)$, the quality of approximation is much worse, the green and yellow zones are limited to 16% and 40% of the potential well depth (about $v$=2 and 5, respectively). The function $\delta(r)$ has a domed shape with maximum 1100 cm$^{-1}$ in the upper part of the potential well. In the main part of the well (up to ~90%, $v = 17$) the $M1(r)$ model alters the shape of the term to a lesser extent than $M2(r)$. The repulsion branches go above the original term, although in the region of the continuous spectrum they should reach the ordinate axis. Below we describe different variants of their behavior, which can be attributed to the difficulties of the theoretical description of the repulsion, so we will consider only the attraction branches.

The distortion of the approximated vibrational structure is illustrated in the Birge-Sponer coordinates (Fig. 3),

$$\omega(v) \equiv \Delta_1 G = G(v+1) - G(v) \tag{5}$$

here $\omega(v)$ are vibrational quanta [1]. The Birge-Sponer extrapolation was proposed even before Morse's publication as an empirical method for estimating the dissociation energy of diatomic molecules [13,14]. Having discovered from the analysis of experimental data that for many molecules the vibrational frequencies in the lower part of the potential well can be approximated by a linear function of the vibrational number $v$, Birge and Sponer extrapolated the frequency values to zero and obtained a value close to the dissociation energy of the molecule. Obviously, this operation is similar to the construction of the Morse model function $M1(r)$, and, as Lessinger demonstrated in [15], from extrapolation in coordinates $\Delta_1 G = f(v + ½)$ for the Morse function one can obtain the value of $D_e$.

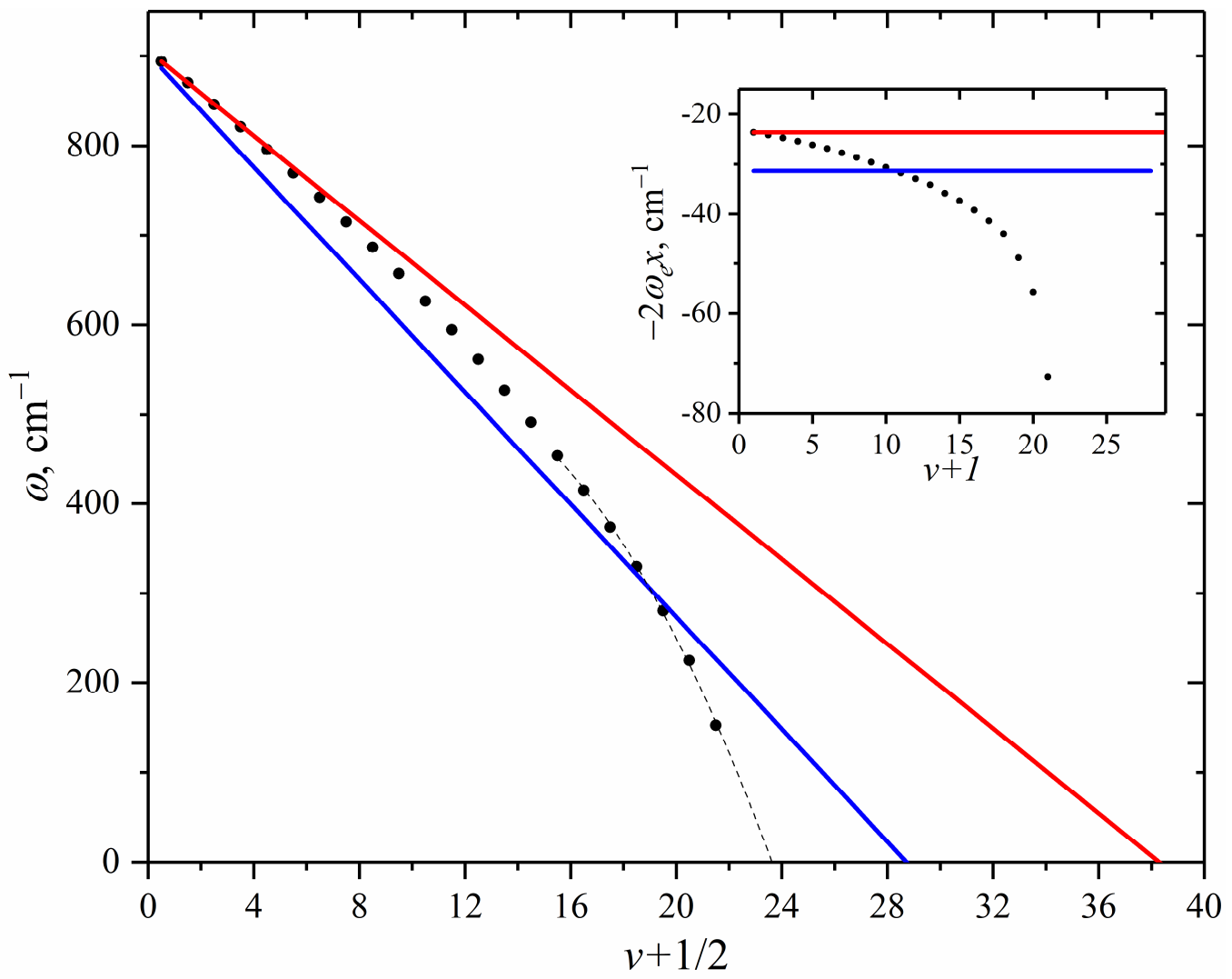

**Figure 3**. Birge-Sponer plot (points) for the term X $^1\Sigma_g^+$ of the $F_2$ molecule according to [9] and its Morse approximations $M1$ (red straight line) and $M2$ (blue straight line). In the inset, the anharmonicity $-2\omega_e x(v)$ is shown (points), with parallel lines illustrating constant values of anharmonicity $-2\omega_e x_e$ for $M1$ (red) and $-2\omega_e x_e'$ for $M2$ (blue).

The straight line M1 (red) in Fig. 3 is drawn through the points $v$ = 1 and 2, which determine the anharmonicity $\omega_e x_e$. It shows the positions of vibrational levels in $M1(r)$, their energies are shown by dashes in Fig. 2 b). Real levels converge to the assymptote faster than for potential $M1(r)$ (the width of the potential well $U(r)$ is larger); the difference between them increases significantly in the upper part of the potential well. In agreement with Eq. (4), the binding energy $D_e$ increases, and fictitious levels emerge in the potential well $M1(r)$, this is also seen in the Birge-Sponer extrapolation. Experimental frequencies (points) constitute a curve (Birge-Sponer plot [14]), extrapolation of which to the abscissa axis by the dashed line, following Gaydon [15], gives the value of the last vibrational level. For the term $M2(r)$, the value of $D_e$ is known, the value of anharmonicity increases, $\omega_e x'_e > \omega_e x_e$ , according to Eq. (4), and the slope of the blue line increases. The vibrational levels converge faster than in $U(r)$ (potential curve $M2(r)$ is wider), and up to the asymptote $D_e$ there are 7 fictitious levels appearing below the asymptote, in agreement with Fig. 2 b).

For our purposes the main advantage of the Birge-Sponer plot is that it gives an initial idea of the changes in the vibrational structure of the term $U(r)$ in the Morse extrapolation. In the case of $F_2$, one can say that for small values of $v$, the frequencies of $M1(r)$, while gradually increasing in comparison with those corresponding for $U(r)$, reproduce the original set quite well, but approximately in the middle of the potential well, the differences between them quickly increase, and the picture becomes uninterpretable. Also, the Birge-Sponer plot reminds us of the existence of fictitious levels.

In addition, let us pay attention to the following circumstance. In Fig.2 a) the $r_2$ values of the vibrational levels of $M2(r)$ from the first level to $v$=20 (blue dashes) are greater than for $U(r)$. Only for levels $v$=21 and 22 these values are smaller. At the same time, Fig. 2 b) shows that the energies of the $M2(r)$ levels decrease as compared to the corresponding values of $U(r)$ due to an increased value of dimensionless anharmonicity $x$ for the $M2(r)$ term. This phenomenon is of a general nature; it occurs for all terms considered below, except for terms with anomalies, when these curves intersect. This is because the $M2(r)$ curve lies below $U(r)$, and the levels of $M2(r)$ with lower energies may have turning points further from zero than $U(r)$ levels.

It is appropriate here to make the following important note. The anharmonicity constant $\omega_e x_e$ is included in the set of basic spectroscopic constants, for example in the reference books [16,17] and often in the current publications of new data. At the same time, the method of calculating of anharmonicity constant for a given term usually remains unclear, - it is calculated either using positions of the first three (or averaged over more than three) levels in the potential well, or as the coefficient of the quadratic term in the approximation of a set of vibrational frequencies by a polynomial (taking into account the zero-point energy). These values are not the same. The constants obtained by any of these approaches get into reference books and sometimes are used to calculate the bond energy $D'_e$ using Eq. (4), which is equivalent to constructing the $M1(r)$ Morse model. Tuttle et al. [18] estimated the obtained values of $D'_e$ for a number of complexes of the same type and compared those with the literature data. The differences in both directions are often several tens of percents. In addition, a small random error in determining the first two frequencies can lead to completely implausible magnitudes of bond energies due to the distant extrapolation. Below we will always determine $\omega_e$ and $\omega_e x_e$ from the first three levels $v$=0,1,2.

For the future, it turns out to be useful to introduce a new concept, – the empirical anharmonicity function (second differences) $-2\omega_e x(v)$, cm$^{-1}$, or $\omega_e x(v)$, cm$^{-1}$, or even dimensionless $x(v)$, [1–4]:

$$\Delta_2 G(v) = \omega(v+1) - \omega(v) \equiv -2\omega_e x(v), \qquad (6)$$

It is formally defined as a change in the frequency of the vibrational quantum when $v$ increases by one, technically - as the second differences in the succession of levels of the vibrational energy of the electronic term $U(r)$, Eq. (6). It can be considered as a generalization of the concept of anharmonicity to the entire potential curve below the asymptote, i.e. the measure of deviation from the parabolic shape of a section of the dependence in Eq. (2) in the range from $v$ to $v + 2$. The value of $-2\omega_e x(v)$ is determined by the positions of three neighboring levels and assigned to the median one. The peculiarity of the anharmonicity function is that the segments $\Delta_2 G(v)$ are not the same, and the position of the level is not in the center of the segment. The shape of this function determines position of the vibrational levels in the potential well and can serve as an individual characteristic of the electronic term, in particular, for the Morse potential $-2\omega_e x(v)$=Const, which characterizes the rate of convergence of vibrational levels. A monotonic increase of anharmonicity function describes the convergence of levels faster than the linear approach of Morse, while areas with decreasing anharmonicity indicate local deviations from the Morse potential. However, attempts to establish a quantitative relationship between the anharmonicity function and the shape of the term are not promising, especially considering the part near the asymptote, where the precision of calculated terms requires experimental verification.

The points in the inset in Fig.3 show the shape of the function $-2\omega_e x(v)$, for the ground term of $F_2$. It increases monotonously with acceleration from 23 to 74 cm$^{-1}$, i.e. 3.2 times higher. The horizontal lines show the values of anharmonicity for $M1(r)$ (red) and $M2(r)$ (blue), which for the Morse potential are constant, −23.6 cm$^{-1}$ and −31.4 cm$^{-1}$ respectively. The length of the lines characterizes the total number of levels, including fictitious ones.

## 3. Results

Using the data from literature, we constructed potential curves of a number of diatomic molecules $U(r)$ and the set of vibrational levels $G(v)$, and calculated a set of vibrational quanta $\Delta_1 G(v)$ and their differences $\Delta_2 G(v)$. Model potential functions $M1(r)$ and $M2(r)$ are constructed, by approximating the term $U(r)$ using the Morse formula, as well as differences $\delta(r) = U(r) - M(r)$, which show the deviations of functions $M1(r)$ and $M2(r)$ from $U(r)$ in these approximations. The changes in the vibrational structure during approximation are described by the function $\omega_e x(v)$, a rough estimate of them can be obtained from the Birge-Sponer diagram $\Delta_1 G(v + ½)$. These results allow for the preliminary classification of the real terms of diatomic molecules with a valence bond into three groups: simple terms with minimal distortion of the contour shape and an increase in the anharmonicity function, which characterizes the changes in the vibrational structure due to approximation, terms with anomalies in the lower part of the potential well, and terms with anomalies near the asymptote due to a change in the type of interatomic interactions at large distances (the deviations of the term from the simple shape will be termed anomalies or peculiarities). Here, it is appropriate to add that the shapes of the function $\omega_e x(v)$ depend on the distances between vibrational levels and can change noticeably, in particular, when switching to isotopic forms of the molecule. In this communication we will consider simple terms, and the terms with anomalies will be analyzed later in the following paper.

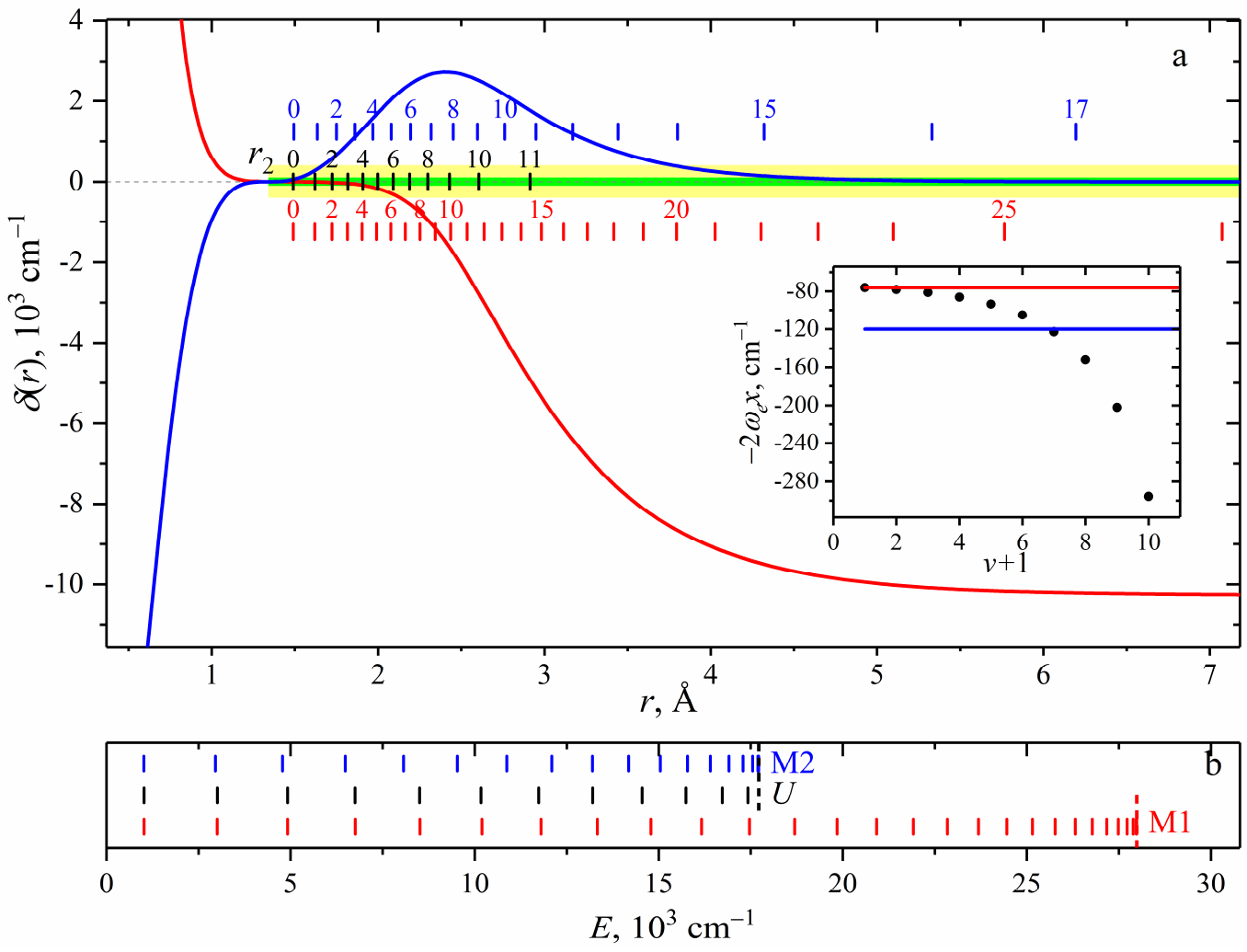

**Figure 4**. a. Deviations of $M1$ (red line) and $M2$ (blue) from the term X $^2\Sigma^+$ of the BeH molecule according to [19]. Vertical dashes near the level of $\delta(r) = 0$, shown by the dashed line, are the values of the outer classical turning points $r_2$ of vibrational levels for potential curves $U(r)$ (black), $M1(r)$ (red), and $M2(r)$ (blue). The abscissa scale of $r_2$ coincides with the abscissa scale of the function $\delta(r)$. In the inset, the anharmonicity $-2\omega_e x(v)$ is shown (points), with parallel lines illustrating constant values of anharmonicity $-2\omega_e x_e$ for $M1$ (red) and $-2\omega_e x_e'$ for $M2$ (blue). b. The energies of vibrational levels respect to the minima of the potential for $U$ (black), $M1$ (red), and $M2$ (blue).

**3.1. Simple terms.** These are the terms with shapes not much different from the Morse contour, and curves $M1(r)$ and $M2(r)$ do not intersect with $U(r)$. The main features of the shape of simple terms are monotonous increase of deviation $\delta(r)$ for $M1(r)$, and the presence of a smooth dome in $\delta(r)$ for $M2(r)$. The anharmonicity function $\omega_e x(v)$ of simple terms increases also monotonously, but often this dependence is broken by the appearance of an abrupt decrease in its value near the asymptote by several times. This feature does not appear in the shape of the functions $U(r)$, $M1(r)$ and $M2(r)$ and requires special theoretical and experimental analysis. The term of the ground state X $^1\Sigma_g^+$ of the $F_2$ molecule, described in detail above, belongs to this group. Here we will consider the term X $^2\Sigma^+$ of BeH radical, as well as the terms of molecules $Li_2$ X $^1\Sigma_g^+$ and $B_2$ X $^3\Sigma_g^-$. In the last two examples, the shapes of simple terms are somewhat distorted by a sharp changes of anharmonicity in the upper part of the potential well, at 2-5% of its depth, which will be considered in analysis of $Li_2$ properties.

**BeH.** The term X $^2\Sigma^+$, together with the one for $F_2$, is a rare example of the most simple Morse approximation. Fig. 4 shows the deviations $\delta(r)$ of the curves $M1(r)$ and $M2(r)$ from the real term of BeH (the ordering of the curves is the same as in Fig. 2 for $F_2$). The functions $M1(r)$ and $M2(r)$ are constructed using the calculated data [19]. The inset shows the values of anharmonicity, they increase from 80 to 295 cm$^{-1}$, i.e. by 3.7 times. The $M1(r)$ curve reproduces the original term in the lower part of the potential with deviation not exceeding 100 $cm^{-1}$ up to $v$~4, 48%, and 400 $cm^{-1}$ up to $v$~6, 66% of the well depth (Fig.5).

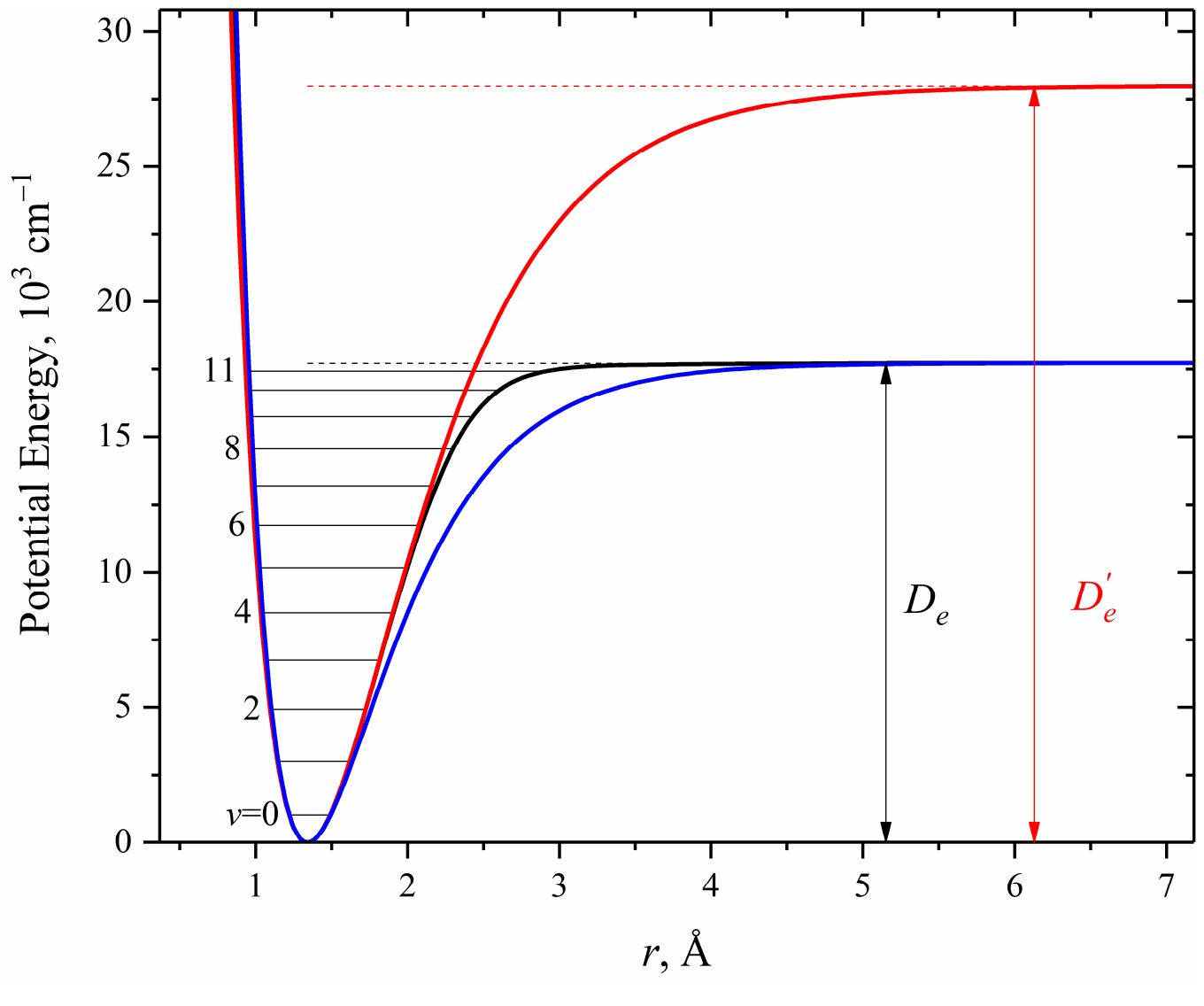

**Figure 5**. The term X $^2\Sigma^+$ of the BeH molecule (black line) according to [19] and its Morse approximations $M1(r)$ (red) and $M2(r)$ (blue). $D_e$ = 17722 cm$^{-1}$, $D_e'$ = 27988 cm$^{-1}$. There are 16 fictitious levels for $M1(r)$ and 6 for $M2(r)$.

Then the attraction branch goes up, the deviation from $U(r)$ increases rapidly, and the $M1(r)$ asymptote exceeds the value of $D_e$ by 10270 $cm^{-1}$, i.e. by 58%. This is the largest deviation value of all terms we have considered. In the resulting potential well, there are 16 fictitious vibrational levels above $D_e$. The $M2(r)$ curve for the attraction branch lies below $U(r)$, its deviation increases, reaching a maximum value of ~3000 $cm^{-1}$ in the upper part of the potential well, at about $v = 9$, and then approaches the expected position near the asymptote (Fig.4). From $v = 9$ to the asymptote (12% of the potential well depth) there are two levels for $U(r)$. The initial function $U(r)$ is the first to reach the asymptote, followed by $M2(r)$ and finally by $M1(r)$. The density of levels in $M2(r)$ is higher than in $U(r)$, so that 4 fictitious vibrational levels appear below asymptote. A Birge-Sponer plot looks similar to the Fig.3, indicating significant overestimation of $D_e'$ and many fictitious vibrational levels, Fig.S1.

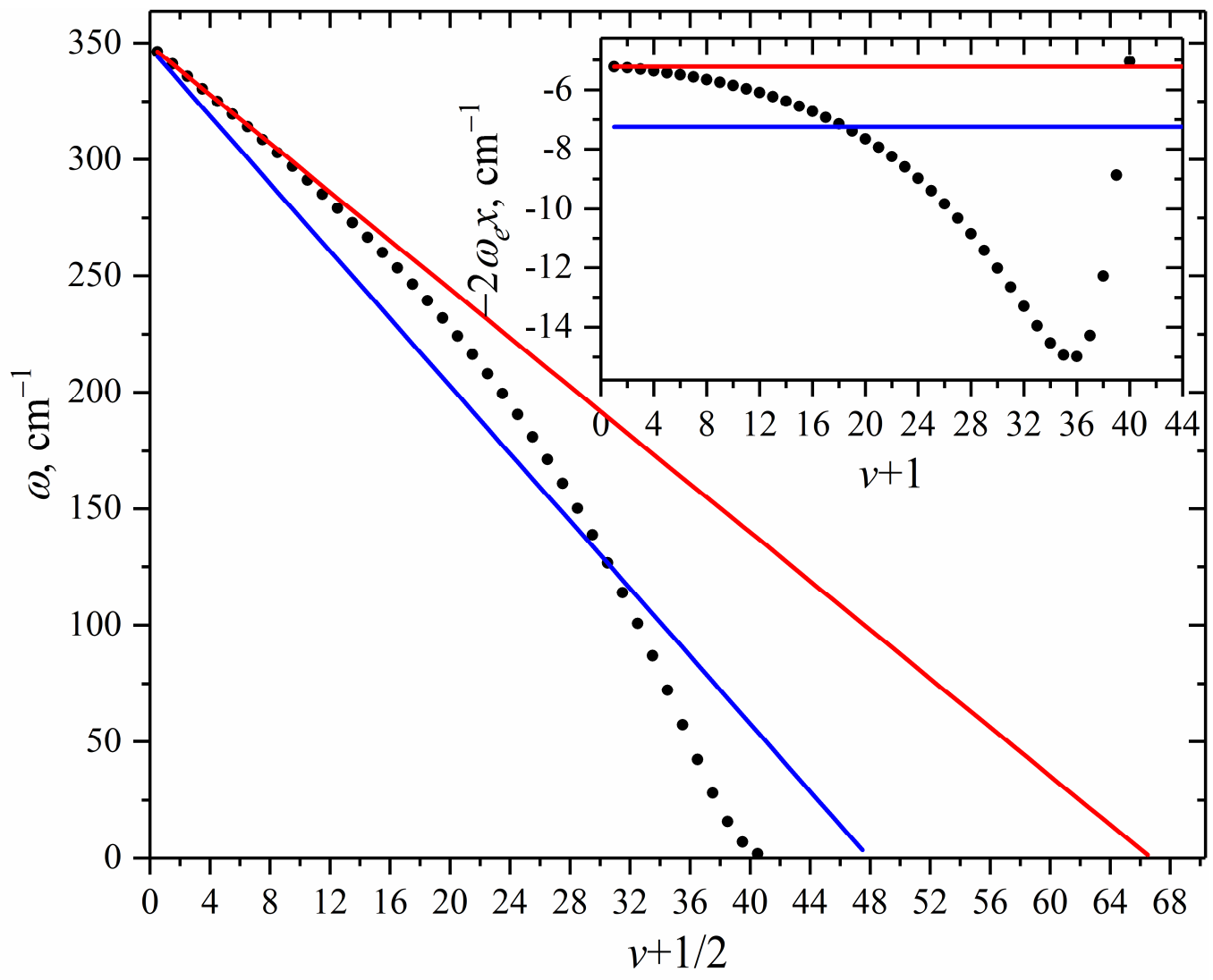

**Figure 6**. Birge-Sponer plot (points) for the term X $^1\Sigma_g^+$ of the $Li_2$ molecule according to [20] and its Morse approximations *M*1 (red straight line) and *M*2 (blue straight line). In the inset, the anharmonicity $-2\omega_e x(v)$ is shown (points), with parallel lines illustrating constant values of anharmonicity $-2\omega_e x_e$ for *M*1 (red) and $-2\omega_e x_e'$ for *M*2 (blue).

**$Li_2$.** Fig. 6 shows the Birge-Sponer diagram and the anharmonicity function $\omega_e x(v)$ for the ground state electronic term X $^1\Sigma_g^+$of the $Li_2$ molecule, constructed using experimental data from [20].

The dependencies are similar to those for $F_2$ and BeH shown above, with the one exception, - monotonous growth of the absolute value of anharmonicity followed by a sharp decrease from 15 to 5.3 $cm^{-1}$ in the range $v$ = 36 – 41, which appears in the Birge-Sponer diagram as a section with negative curvature (concavity) and a sharp bend near the asymptote. This range corresponds to ~1% of the depth of the potential well below the asymptote. The decrease in anharmonicity in excited vibrational states was first noted in [1] when analyzing calculations of the potential curves of the main terms for $Li_2$ in [20] and for $O_2$ in [21]. The authors [1] noticed the similarity with the anharmonicity of van der Waals molecules [22], in which it decreases throughout the sequence of all vibrational levels, and also noted the coincidence of calculated frequencies near the asymptote with the experimental values in [23]. In [23] the vibrational-rotational spectra of $Li_2$ dimers in the upper half of the potential well are approximated by the potential in the form of an inverse power series in $r$. The verification of the results for the last values of $v$ was carried out using the Le Roy formula [24] (cited in [23]) according to which the dependence of the rotational constant $B_v(v)$ for the highest $v$ looks like a straight line, crossing abscissa near $v_{max}$. In Fig. 6 in [23] this dependence describes well the available experimental data for $v \geq 35$, but deviations are observed when $v$ decreases from $v$ = 34. Essentially, this means the dominance of van der Waals interactions, but the authors do not use this term, as well as the concept of anharmonicity. We have constructed the same graph using experimental data from [25], which clearly shows the deviation from the straight line of Le Roy, starting at $v$ ~ 36 – 35, Fig.S3 in Supp. Mat. Thus, the kink in anharmonicity near the asymptote in the vibrational structure of $Li_2$ and several other molecules can be explained by a change in the nature of interatomic interaction at high values of the external turning point $r_2$. It is important to emphasize that in the same part of the potential there is a jump in the anharmonicity function, starting at $v$ = 35, which is a strong confirmation of the assumption about the nature of this jump made in [1]. An additional argument in favor of such interpretation of the cause of the detected anomaly is the behavior of anharmonicity in the terms of van der Waals molecules - it decreases monotonically for ArXe, $Xe_2$, $Kr_2$ over the entire range of vibrational frequencies [1].

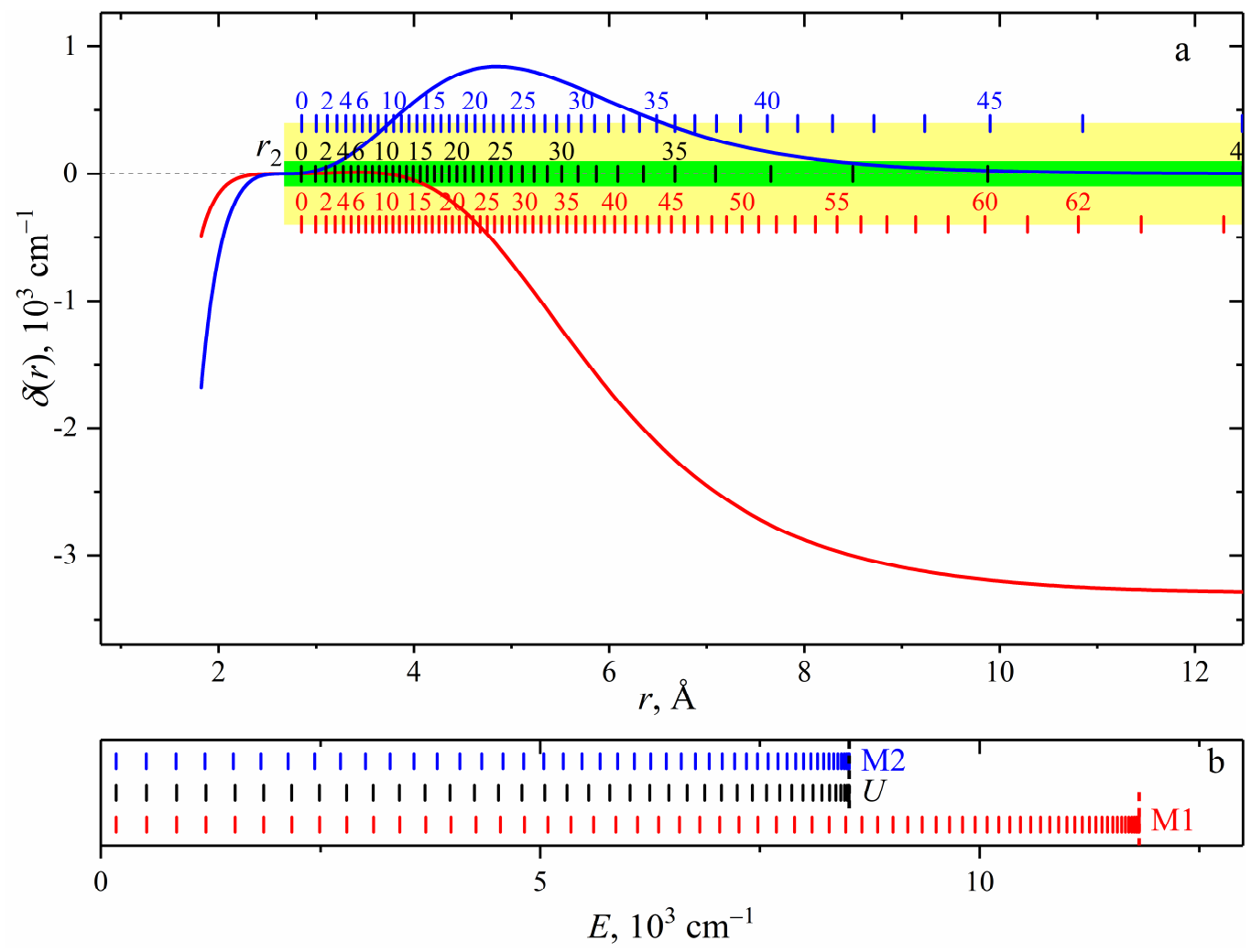

**Figure 7**. a. Deviations of $M1(r)$ (red line) and $M2(r)$ (blue) from the term X $^1\Sigma_g^+$ of the $Li_2$ molecule according to [25]. Vertical dashes near the level of $\delta(r) = 0$, shown by the dashed line, are the values of the outer classical turning points $r_2$ of vibrational levels for potential curves $U(r)$ (black), $M1(r)$ (red), and $M2(r)$ (blue). The abscissa scale of $r_2$ coincides with the abscissa scale of the function $\delta(r)$. **b**. The energies of vibrational levels respect to the minima of the potential for $U$ (black), $M1$ (red), and $M2$ (blue).

One can try to use the van der Waals radius of atoms for comparison, which for Li is equal to $R_{vdW}$=1.82 Å. At the anharmonicity maximum at $v$ = 36, the value of $r$ = 7.1 Å is almost 2 times higher than $2R_{vdW}$ = 3.64 Å. Apparently, this indicates the possibility of using anharmonicity to probe for van der Waals interactions.

Fig. 7 shows the deviations of $M1(r)$ and $M2(r)$ from the real potential curve for $Li_2$ X $^1\Sigma_g^+$ that occur with approximation using the Morse formula. The difference curve $M1(r)$ monotonically deviates from the real term towards higher energies and reaches an asymptote at ~3300 cm$^{-1}$, which is 38% greater than the true value $D_e$=8517 cm$^{-1}$, see also the plot of these terms in Fig. S3 in Supp. Mat. The largest deviation for the $M2(r)$ from the real term, 850 cm$^{-1}$, shows a maximum around $v$ =25 at ~1500 cm$^{-1}$ from the asymptote (17% of the potential well depth). The range of precision better than 100 cm$^{-1}$ and the range of error within 400 cm$^{-1}$ are shown in Fig. 7 by color.

On deviation curves for $M1(r)$ and $M2(r)$ there are no visible indications of anomalies of the potential curve of the real term near the asymptote. There are 27 and 8 ficticious vibrational levels in $M1(r)$ and $M2(r)$, similar results can be seen in Fig. 6.

**$B_2$.** Presented in Fig. 8, the deviations $\delta(r)$ and anharmonicity $-2\,\omega_e x(v)$ for the main term X $^3\Sigma_g^-$ of the molecule $^{11}B_2$ according to the theoretical data of [26] are also typical for a simple term, except for a sharp decrease in anharmonicity near the asymptote, which is similar to $Li_2$. Curve $M1(r)$ reproduces $U(r)$ better than $M2(r)$ in the range up to $r$ = 2.62 Å, near $v$ = 24, this is about 83% of the potential well. At this point, the deviation from the initial term is about 1800 cm$^{-1}$, after that $M2(r)$ becomes more preferable. The last level for $^{11}B_2$ is $v_{max}$= 38, there are 22 fictitious levels for $M1(\mathrm{r})$ 22 and 7 for $M2(\mathrm{r})$ . The extrapolated value of the bond energy $D_e'$ is ~31500 cm$^{-1}$, which is 33% higher than the real value of 23686.30 cm$^{-1}$ (see the plot for $U(r)$, $M1(r)$ and $M2(r)$, Fig.S4 in Supp. Mat.). The deviation of $M1(r)$ from the real term does not exceed 100 cm$^{-1}$ up to $r$ = 2.04 Å (near $v$ = 9), and 400 cm$^{-1}$ up to 2.23 Å (near $v$ = 14) corresponding to ~38% and 55% of the potential well. The maximum deviation of $M2(r)$ from the real term, 1850 cm$^{-1}$, is observed at the upper part of the potential well, near $v$ = 24.

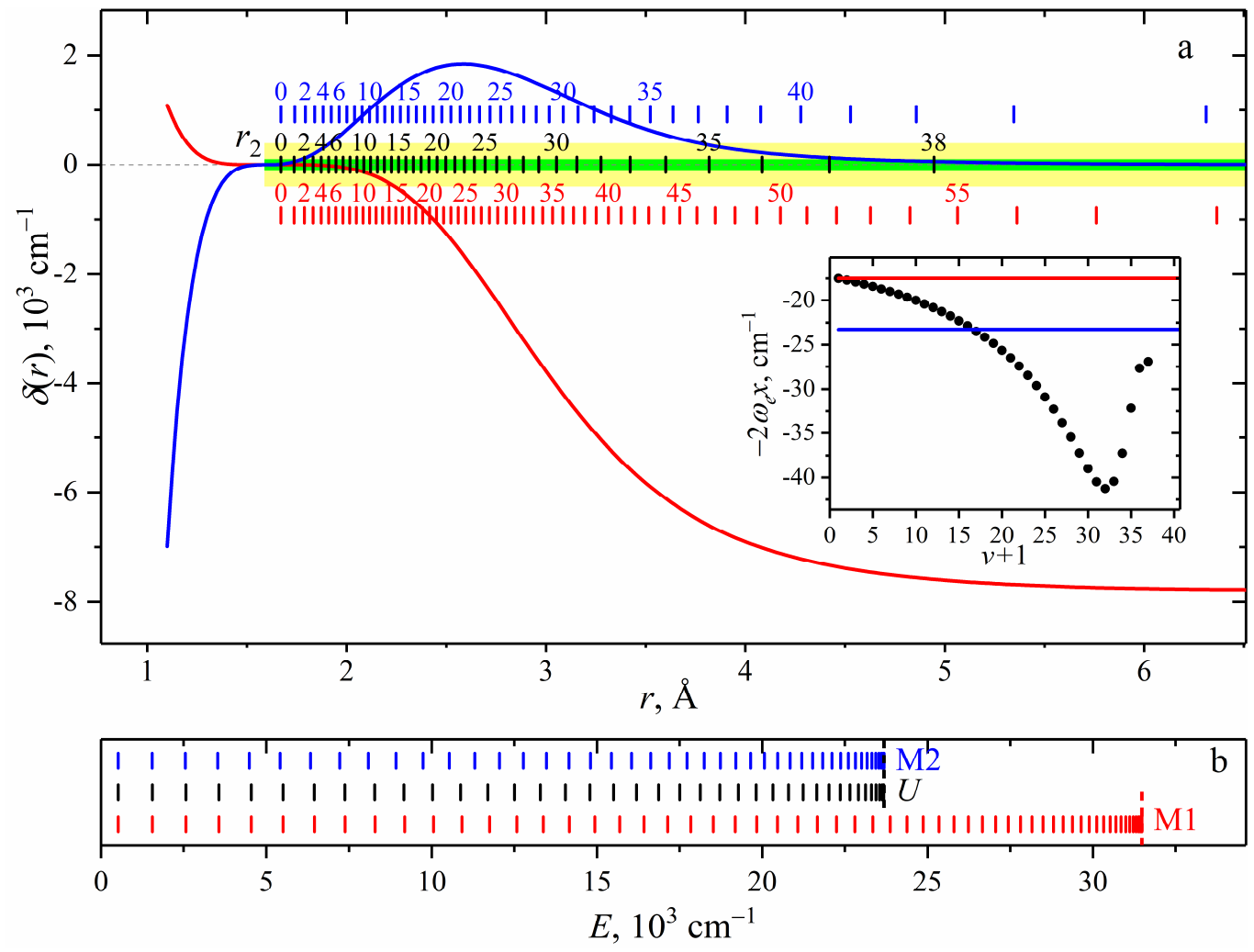

**Figure 8**. a. Deviations of M1 (red line) and M2 (blue) from the term X $^3\Sigma_g^-$ of the $B_2$ molecule according to [26]. Vertical dashes near the level of $\delta(r) = 0$, shown by the dashed line, are the values of the outer classical turning points $r_2$ of vibrational levels for potential curves $U(r)$ (black), $M1(r)$ (red), and $M2(r)$ (blue). The abscissa scale of $r_2$ coincides with the abscissa scale of the function $\delta(r)$. In the inset, the anharmonicity $-2\omega_e x(v)$ is shown (points), with parallel lines illustrating constant values of anharmonicity $-2\omega_e x_e$ for $M1$ (red) and $-2\omega_e x_e'$ for $M2$ (blue). b. The energies of vibrational levels respect to the minima of the potential for $U$ (black), $M1$ (red), and $M2$ (blue).

The jump at $v$ = 32 corresponds to a section of the potential well below the asymptote by 774 cm$^{-1}$ (3.3%), where the van der Waals attractions prevail in the interatomic interaction for the molecules in highly excited vibrational states. In the states with $v$ > 32, van der Waals attraction dominates at values $r$ > 3.28 Å. The van der Waals radius of the B atom is 1.92 Å, therefore the inter-nuclear distance of the van der Waals molecule, at which the interaction could noticeably affect the vibrational structure of 3.84 Å, requires an adequate theoretical analysis.

The Birge-Sponer diagram for $B_2$ shown in Fig.9 is similar to the $Li_2$ diagram - the inflection point at about $v$ = 33 between the convex lower part of the well and the concave upper part, formed by a characteristic bend near the asymptote. The numbers of fictitious levels obtained using Birge-Sponer approximation, 20 and 5, is slightly different from the precise results in Fig. 8b.

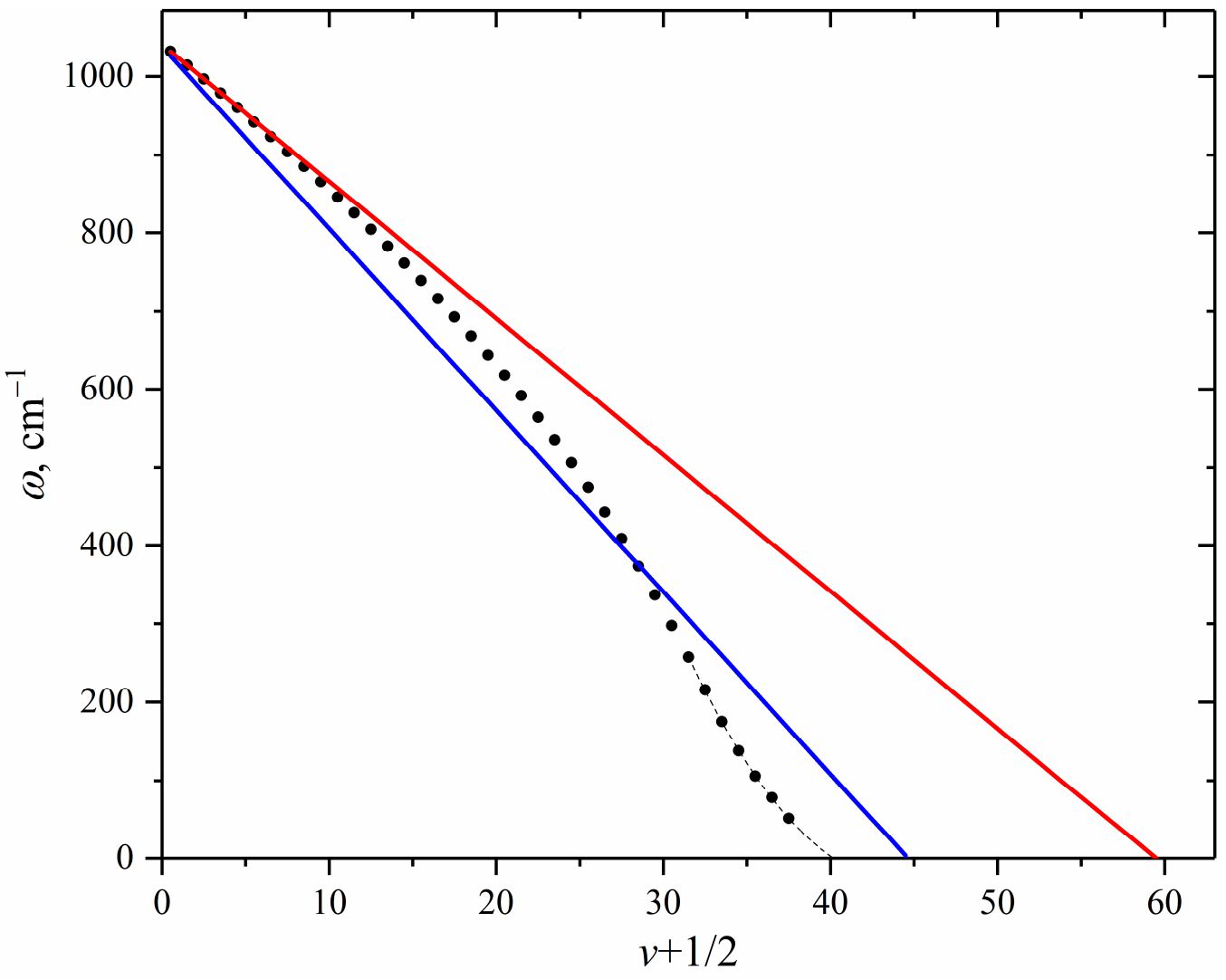

**Figure 9**. Birge-Sponer plot (points) for the term X $^3\Sigma_g^-$ of the $B_2$ molecule according to [26] and its Morse approximations $M1$ (red straight line) and $M2$ (blue straight line).

The potential curves $U(r)$, $M1(r)$ and $M2(r)$ are shown in Fig.S4 in Supp. Mat.

## 4. Discussion

Morse's task was to create a potential function with a single anharmonicity coefficient $\omega_e x_e$ (more precisely, a dimensionless coefficient $x_e$) for diatomic molecules, which would allow obtaining a solution to the Schrödinger equation in an analytical form. Such a function could be the next approximation in molecular spectroscopy after the harmonic oscillator. This approximation is rough due to the linear convergence of vibrational levels to the asymptote, but it would facilitate the semi-quantitative solution of many problems of applied spectroscopy.

The analysis of theoretical and experimental literature data on potential curves and their vibrational levels allows deducing some properties of the Morse function and the results of its application to approximation of the real terms.

The Morse equation provides a solution to two types of problems – 1) estimation of the binding energy $D_e$ using a given value of anharmonicity $\omega_e x_e$, and 2) construction of the potential energy curve using a known binding energy, which allows one to find the anharmonicity coefficient. These two alternative parameters are included in the Morse formula and cannot be used simultaneously, since they are related by the Eq. (4), which is valid only for the Morse potential. Eq. (4) connects these two parameters and makes it possible to obtain any of the two solutions to the Morse problem, the functions $M1(r)$ and $M2(r)$, respectively. The ambiguity of the approximation of the real term $U(r)$ by the Morse formula is discussed by Urbanczyk and coauthors [7,27,28].

For construction of the $M1(r)$ function, the values of the harmonic frequency $\omega_e$ and the anharmonicity $\omega_e x_e$, (or $x_e$) are used, which are usually determined by two experimental vibrational frequencies in the lowest part of the potential well. In real terms of molecules with a valence bond, vibrational levels usually converge faster the linear law, therefore, for $M1(r)$ with underestimated anharmonicity, a distant extrapolation to the asymptote gives energy values about 10-50% higher than the true ones. The $M1(r)$ curve approximates well the real term in the lower part of the potential well, the deviations towards higher energies increase monotonously with increasing $v$, the upper part of the potential curve lies in the region of a continuous spectrum. The total number of calculated vibrational levels increases, but their number within the potential well $U(r)$ usually decreases.

The $M2(r)$ approximation uses the $D_e$ value known from an independent source, the asymptotes of the curves $M2(r)$ and $U(r)$ coincide, therefore the anharmonicity coefficient turns out to be higher than that of $M1(r)$. The harmonic frequency $\omega_e$ is the same as for M1, and the change in anharmonicity emerges due to an increase in the dimensionless coefficient $x_e$. For the M2 approximation, the density of vibrational levels increases, and positions of even the lowest levels are lower than for the $U(r)$. The number of levels also increases, the extra levels are adjacent to the asymptote. The non-monotonous deviation of $M2(r)$ from $U(r)$ towards lower energies is bell-shaped, it is small near the asymptote and at the bottom of the potential well, and reaches maximum in the upper half of the well. In general, the $M1(r)$ curve approximates simple terms $U(r)$ more accurately than $M2(r)$. Both functions give an overestimated number of vibrational levels. The excess levels in $M1(r)$ are above the asymptote of the real term, in $M2(r)$ they are below the asymptote.

The features of the shape of the term $U(r)$ are revealed by the difference function $\delta(r) = U(r) - M(r)$, which characterizes the deviation of the real contour from the simple Morse approximation with constant anharmonicity. A novel approach was suggested in the work by McCoy [29], see also [1].

The deviations of the vibrational structure from the linear dependence on $v$ can be estimated by comparing the first differences $\Delta_1 G(v)$, Eq. (5), for the real term and the approximating functions. The $\Delta_1 G(v)$ values represent a set of vibrational quanta of a given electronic state, and for the Morse function they converge following a linear law. A comparison with the Birge-Sponer extrapolation, which is almost identical to the $M1(r)$ model, shows that in $\Delta_1 G(v + ½)$ coordinates it illustrates the vibrational structure of the $M1(r)$ and $M2(r)$ models. A more accurate assessment of these deviations is made using the anharmonicity function, defined as the second difference $\Delta_2 G(v) \equiv -2\omega_e x(v)$, or $\omega_e x(v)$, Eq. 6. It demonstrates the anharmonicity within a section of a potential well in an interval of three vibrational quantum numbers ($v$−1, $v$, $v$+1) and can serve as an individual characteristic of the term. For the Morse function it is a constant. The changes in positions of vibrational levels $v$ caused by an approximation are determined by the differences between the values of the anharmonicity function of the terms $U(r)$ and $M1(r)$ or $M2(r)$. The positions of the vibrational levels of these terms on a given potential curve are given by the coordinates $r_2$, the outer classical turning points.

Based on the form of the functions $\delta(r)$ and $\omega_e x(v)$, we identified terms $U(r)$ with small deviations from the Morse formula (simple terms), terms with distortions in the lower part of the potential well, terms with distortions in the upper part of the well, and terms of molecules with van der Waals bonds, which have special properties [1–3]. Simple terms are distinguished by the monotonous increase of their anharmonicity and the modulus of the function $\delta(r)$ for $M1(r)$, with the term $U(r)$ located between the curves $M1(r)$ and $M2(r)$. A common feature of the behavior of the anharmonicity functions for $U(r)$ that is not reflected in the shape of the contour and models is a sharp decrease in anharmonicity near the asymptote. There are certain reasons to attribute such effects to changes in the type of interatomic interactions from valence to van der Waals upon significant stretching of the molecule in highly excited vibrational states.

This abrupt transition is not always observed even for molecules with weak bonds and at a large internuclear distance. On the other hand, for molecules with van der Waals bonds, anharmonicity increases over the entire energy range [1]. Finally, there are examples of slow increase of anharmonicity a great distance from the asymptote [1,4]. Apparently, there is a dependence on the electronic state of the atoms of the dissociation products, and at this stage an empirical classification of the observed states can be useful. At this stage, it can be assumed that a sharp decrease in anharmonicity at a distance of 1-3% of the asymptote energy can also occur for simple terms.

Formally, the main systematic error of the Morse approximation is a result of the formulation of the problem, – the use of a constant value of the anharmonicity of $\omega_e x_e$ and of the Eq. (4) connecting $\omega_e x_e$ and $D_e$. The possible *a priori* estimate of the error introduced by Eq. (4) is determined by the degree of its relevance. For $M1(r)$ approximation, this equation is used to calculate the binding energy $D_e$ based on the value of anharmonicity estimated from the first two vibrational frequencies. In the $M2(r)$ model, on the contrary, the exact value of $D_e$ is used to obtain the value of $\omega_e x_e'$ averaged over all vibrational frequencies. Preliminary attempts to estimate the changes in $M1(r)$ and $M2(r)$ due to the averaging of anharmonicity over several levels are made in the works [1,3]. For the random errors, it is important to take into account the accuracy of the experimental determination of the basic parameters of $\omega_e$ and $\omega_e x_e$, the two first vibrational transition frequencies $\omega(0-1)$ and $\omega(0-2)$. The far extrapolation can significantly distort the results.

## 5. Conclusion

In conclusion, we analyzed the two alternative approximations of the real potential $U(r)$ of diatomic molecules by the Morse potential with a constant value of anharmonicity. The difference function $\delta(r) = U(r) - M(r)$ is introduced for classifying the shapes of the real potential wells. The M1 approximation is based on anharmonicity calculated from the first three vibrational levels, and describes better the potential well in the lower part. However, in many cases this approximation overestimates the dissociation energy by 10 – 50%, and therefore generates many fictitious vibrational levels near and above the real asymptote. The $M2(r)$ approximation is based on calculation of the anharmonicity from the true value of dissociation energy, and in most cases provides a worse approximation of the real potential curve. However, it generates less fictitious vibrational levels and provides better approximation in the upper part of potential well. The difference function $\delta(r)$ was used to distinguish between molecules with small deviations at the bottom of potential well (simple terms), terms with deviations in the upper part of the potential well, and terms of van der Waals molecules. The transition from the simple shape of $U(r)$ to the van der Waals behavior in the upper part of potential well can be detected using the difference function $\delta(r)$, as illustrated for $Li_2$ and $Be_2$ molecules.

## Acknowledgments

The authors are grateful to Dr. I.G. Denisov for systematic creative consultations, Professors V.P. Bulychev, H.-H. Limbach, I.G. Shenderovich, who read the article and made important comments.

Supplementary materials for

# Approximation of electronic term of diatomic molecule by the Morse function. The role of anharmonicity. I. Simple terms.

by

R.E. Asfin*, G.S. Denisov

*Department of Physics, Saint Petersburg State University, 7/9 Universitetskaya Nab., 199034 Saint Petersburg, Russian Federation*

*Corresponding author: R.Asfin@spbu.ru

Both $M1(r)$ and $M2(r)$ functions were built according the Morse function:

$$M(r) = D_e\left[1 - \exp\left(-a(r - r_e)\right)\right]^2$$

The parameter $r_e$ is the equilibrium distance of the approximated function $U(r)$ measured in Å and was taken from the description of this function. The parameter $D_e$ [cm$^{-1}$] for $M1(r)$ was calculated according the Eq. (4) of the article. For $M1(r)$ we denoted it as $D_e'$:

$$D_e' = \omega_e^2/4\omega_e x_e$$

The harmonic wavenumber $\omega_e$ and anharmonicity $\omega_e x_e$ (both in cm$^{-1}$) were obtained from the wavenumbers of the fundamental $\nu_{10}$ and the first overtone $\nu_{20}$ transitions. The values these magnitudes were taken from the articles, where $U(r)$ is described, or calculated from the energy of the first three vibrational levels.

$$\omega_e x_e = -(\nu_{20} - 2\nu_{10})/2$$

$$\omega_e = \nu_{10} + 2\omega_e x_e$$

It should be mentioned that the $\omega_e$ defined by such equation may not coincide with the harmonic wavenumber obtained via the second derivative of $U(r)$ in $r_e$ and the reduced mass of the molecule.

For $M2(r)$ the $D_e$ equal to the dissociation energy of $U(r)$, measured from the bottom of the potential well, and anharmonicity $\omega_e x_e'$ was calculated as

$$\omega_e x_e' = \omega_e^2/4D_e$$

$\omega_e$ is the same as for $M1(r)$.

The exponent $a$ [Å$^{-1}$] was obtained from

$$a = 2\pi \frac{\omega_e}{\sqrt{2\frac{hD_e}{\mu Mc}}} 10^{-8}$$

where $h$ = 6.62607015 $10^{-27}$ [erg s] is the Planck constant, $\mu$ [Da] is reduce mass of the molecule, $M$ = 1.660539 $10^{-24}$ [g/Da] is atomic mass constant, $c$ = 29979245800 [cm/s] is the speed of light. Coefficient $10^{-8}$ converts cm$^{-1}$ to Å$^{-1}$.

The points in the figures are connected with splines. The splines were used also for calculation of classical turning points.

Below we presented the issues of required values and calculated parameters of the Morse potential $M1(r)$ and $M2(r)$, the vibrational levels for $U(r)$, its function of anharmonicity $-2\omega_e x$, calculated according Eq. (6), and for $M1(r)$ and $M2(r)$ (Eq. (2)) of the main article.

**$F_2$**. The potential $U(r)$ of X $^1\Sigma_g^+$ state of the $F_2$ molecule was built with the analytical even-tempered Gaussian expansion Eq.(1b) [1] with parameters collected in Table I in the line "Empirical". The composite potential energy curve W(R) (Eq. 3a) was also taken into account. The energies of the vibrational levels were obtained from Table II, column "Expt." and shifted to the value of the zero point energy from this Table. The parameters $D_e$ and $r_e$ are also taken from this Table. The calculated parameters of Morse potentials are tabulated in **Table S1**. The vibrational energies of levels $v$ for potentials $U(r)$, taken from [1], $M1(r)$ and $M2(r)$ and the function of anharmonicity are presented in **Table S2**.

**Table S1**. The parameters of the Morse potentials $M1$ and $M2$ for the X $^1\Sigma_g^+$ state of the $F_2$ molecule

| Parameter | $M1$ | $M2$ |
|---|---|---|
| $\mu$, Da | 9.499201581 | |
| $r_e$, Å | 1.41193 | |
| $D_e$,[a] $cm^{-1}$ | 17799.11 | 13408.49 |
| $\omega_e$ , $cm^{-1}$ | 917.55 | |
| $\omega_e x_e$,[b] $cm^{-1}$ | 11.825 | 15.697 |
| $x_e$[c] | 0.01289 | 0.01711 |
| $a$, $Å^{-1}$ | 2.5813 | 2.9741 |

[a] $D_e'$ for M1; [b] $\omega_e x_e'$ for M2; [c] $x_e'$ for M2.

**Table S2**. The energies ($cm^{-1}$) of levels $v$ for potentials $U(r)$ of the X $^1\Sigma_g^+$ state of the $F_2$ molecule [1], and its approximations $M1(r)$ and $M2(r)$. The anharmonicity function $-2\omega_e x$ ($cm^{-1}$) of $U(r)$ is presented in the third column.

| $v$ | $U$ | $-2\omega_e x$ | M1 | M2 |
|---|---|---|---|---|
| 0 | 455.37 | | 455.82 | 454.85 |
| 1 | 1349.27 | −23.65 | 1349.72 | 1341.01 |
| 2 | 2219.52 | −24.18 | 2219.97 | 2195.77 |
| 3 | 3065.59 | −24.76 | 3066.57 | 3019.14 |
| 4 | 3886.9 | −25.41 | 3889.52 | 3811.11 |
| 5 | 4682.8 | −26.14 | 4688.82 | 4571.69 |
| 6 | 5452.56 | −26.90 | 5464.47 | 5300.87 |
| 7 | 6195.42 | −27.74 | 6216.47 | 5998.66 |
| 8 | 6910.54 | −28.66 | 6944.82 | 6665.06 |
| 9 | 7597 | −29.61 | 7649.52 | 7300.06 |
| 10 | 8253.85 | −30.66 | 8330.57 | 7903.67 |
| 11 | 8880.04 | −31.75 | 8987.97 | 8475.88 |
| 12 | 9474.48 | −32.92 | 9621.72 | 9016.70 |
| 13 | 10036 | −34.13 | 10231.82 | 9526.13 |
| 14 | 10563.39 | −35.79 | 10818.27 | 10004.16 |
| 15 | 11054.99 | −37.32 | 11381.07 | 10450.80 |
| 16 | 11509.27 | −39.22 | 11920.22 | 10866.04 |
| 17 | 11924.33 | −41.40 | 12435.72 | 11249.89 |
| 18 | 12297.99 | −44.03 | 12927.57 | 11602.34 |
| 19 | 12627.62 | −48.90 | 13395.77 | 11923.40 |
| 20 | 12908.35 | −55.71 | 13840.32 | 12213.07 |
| 21 | 13133.37 | −72.64 | 14261.22 | 12471.34 |
| 22 | 13285.75 | | 14658.47 | 12698.21 |
| 23 | | | 15032.07 | 12893.70 |
| 24 | | | 15382.02 | 13057.79 |
| 25 | | | 15708.32 | 13190.48 |
| 26 | | | 16010.97 | 13291.78 |
| 27 | | | 16289.97 | 13361.69 |
| 28 | | | 16545.32 | 13400.20 |
| 29 | | | 16777.02 | 13407.32 |
| 30 | | | 16985.07 | |
| 31 | | | 17169.47 | |
| 32 | | | 17330.22 | |
| 33 | | | 17467.32 | |
| 34 | | | 17580.77 | |
| 35 | | | 17670.57 | |
| 36 | | | 17736.72 | |
| 37 | | | 17779.22 | |
| 38 | | | 17798.07 | |

**BeH**. The curve $U(r)$ of X $^2\Sigma^+$ of the $^9$BeH molecule was copied from Table III of the article [2]. The vibrational terms values were taken from Table IV of this article (column "Calc.") with taking into account calculated zero-point energy, given in the footnote of the table. $D_e$ and $r_e$ were obtained from Table II, column "CV+F+R". The parameters used for constructing of $M1(r)$ and $M2(r)$ are presented in **Table S3** and energies of vibrational levels in **Table S4**.

**Table S3**. The parameters of the Morse potentials $M1$ and $M2$ for the term X $^2\Sigma^+$ of the BeH molecule

| Parameter | $M1$ | $M2$ |
|---|---|---|
| $\mu$, Da | 0.906456721 | |
| $r_e$, Å | 1.34099 | |
| $D_e$,[a] cm$^{-1}$ | 27988 | 17722 |
| $\omega_e$ , cm$^{-1}$ | 2062.55 | |
| $\omega_e x_e$,[b] cm$^{-1}$ | 38.00 | 60.01 |
| $x_e$[c] | 0.01842 | 0.02910 |
| $a$, Å$^{-1}$ | 1.4243 | 1.7900 |

[a] $D_e'$ for M1; [b] $\omega_e x_e'$ for M2; [c] $x_e'$ for M2.

**Table S4**. The energies (cm$^{-1}$) of levels $v$ for potentials $U(r)$ of the X $^2\Sigma^+$ state of the $^9$BeH molecule [2], and its approximations $M1(r)$ and $M2(r)$. The anharmonicity function $-2\omega_e x$ (cm$^{-1}$) of $U(r)$ is presented in the third column.

| $v$ | $U$ | $-2\omega_e x$ | $M1$ | $M2$ |
|---|---|---|---|---|
| 0 | 1022.33 | | 1021.78 | 1016.27 |
| 1 | 3008.88 | −76.00 | 3008.33 | 2958.8 |
| 2 | 4919.43 | −78.08 | 4918.88 | 4781.3 |
| 3 | 6751.9 | −81.36 | 6753.43 | 6483.78 |
| 4 | 8503.01 | −86.25 | 8511.98 | 8066.24 |
| 5 | 10167.87 | −93.64 | 10194.53 | 9528.67 |
| 6 | 11739.09 | −105.12 | 11801.08 | 10871.08 |
| 7 | 13205.19 | −122.74 | 13331.63 | 12093.46 |
| 8 | 14548.55 | −152.09 | 14786.18 | 13195.83 |
| 9 | 15739.82 | −202.04 | 16164.73 | 14178.17 |
| 10 | 16729.05 | −295.60 | 17467.28 | 15040.48 |
| 11 | 17422.68 | | 18693.83 | 15782.77 |
| 12 | | | 19844.38 | 16405.04 |
| 13 | | | 20918.93 | 16907.28 |
| 14 | | | 21917.48 | 17289.51 |
| 15 | | | 22840.03 | 17551.7 |
| 16 | | | 23686.58 | 17693.88 |
| 17 | | | 24457.13 | 17716.03 |
| 18 | | | 25151.68 | |
| 19 | | | 25770.23 | |
| 20 | | | 26312.78 | |
| 21 | | | 26779.33 | |
| 22 | | | 27169.88 | |
| 23 | | | 27484.43 | |
| 24 | | | 27722.98 | |
| 25 | | | 27885.53 | |
| 26 | | | 27972.08 | |
| 27 | | | 27982.63 | |

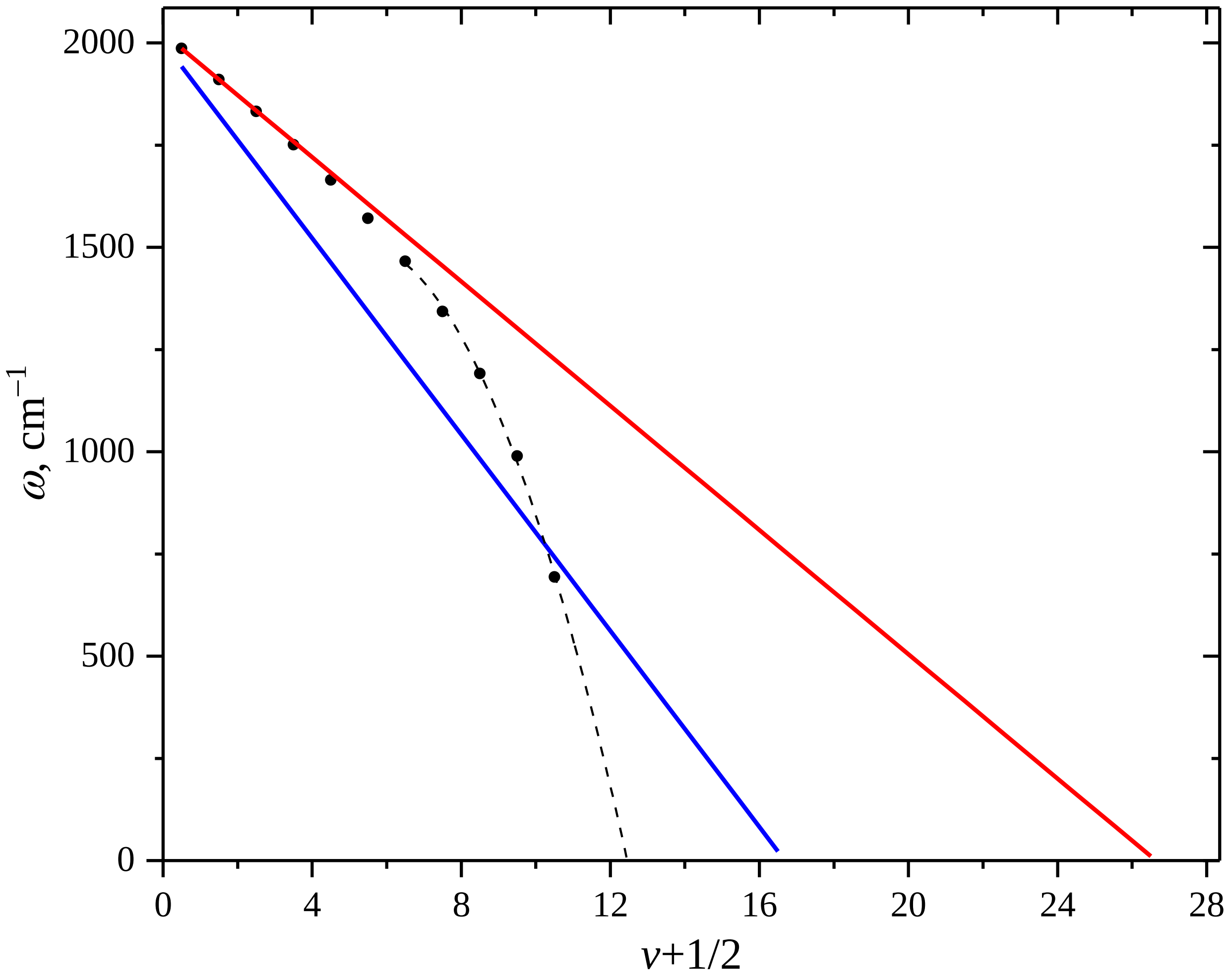

**Figure S1**. Birge-Sponer plot (points) for the term X $^2\Sigma^+$ of the BeH molecule according to [2] and its Morse approximations $M1(r)$ (red straight line) and $M2(r)$ (blue straight line).

**$Li_2$**. Despite the analytical form of $U(r)$ of X $^1\Sigma_g^+$ of the $Li_2$ molecule presented in two articles [3,4] we was not able to build this potential with parameters given in these article. So, we forces to use potential obtained from classical turning points tabulated in Table 6 of the article [4]. The rotational constants $B$ were also taken from this Table. Values of $D_e$ and $r_e$ was found in Table 2 of the mentioned article. The vibrational energies were taken from [3] (SI, file Li2X-constants.txt), because the additional highest level was found in this work. The other levels are close to values in [4]. The parameters used to build $M1(r)$ and $M2(r)$ potentials are presented in **Table S5** and energies of vibrational levels in **Table S6**.

**Table S5**. The parameters of the Morse potentials $M1(r)$ and $M2(r)$ for the X $^1\Sigma_g^+$ term of the $Li_2$ molecule

| Parameter | *M*1 | *M*2 |
|---|---|---|
| $\mu$, Da | 3.508001717 | |
| $r_e$, Å | 2.67299391 | |
| $D_e$,[a] $cm^{-1}$ | 11815.23 | 8516.769 |
| $\omega_e$ , $cm^{-1}$ | 351.449 | |
| $\omega_e x_e$,[b] $cm^{-1}$ | 2.6135 | 3.6257 |
| $x_e$[c] | 0.00744 | 0.01032 |
| $a$, $Å^{-1}$ | 0.73747 | 0.86861 |

[a] $D_e'$ for *M*1; [b] $\omega_e x_e'$ for *M*2; [c] $x_e'$ for *M*2.

**Table S6**. The energies ($cm^{-1}$) of levels $v$ for potentials $U(r)$ of the X $^1\Sigma_g^+$ state of the $Li_2$ molecule [3], and its approximations $M1(r)$ and $M2(r)$. The anharmonicity function $-2\omega_e x$ ($cm^{-1}$) of $U(r)$ is presented in the third column.

| $v$ | $U$ | $-2\omega_e x$ | *M*1 | *M*2 |
|---|---|---|---|---|
| 0 | 175.03 | | 175.07 | 174.82 |
| 1 | 521.25 | −5.23 | 521.29 | 519.02 |
| 2 | 862.24 | −5.26 | 862.29 | 855.96 |
| 3 | 1197.97 | −5.31 | 1198.06 | 1185.66 |
| 4 | 1528.39 | −5.37 | 1528.60 | 1508.10 |
| 5 | 1853.44 | −5.43 | 1853.91 | 1823.29 |
| 6 | 2173.07 | −5.50 | 2174.00 | 2131.23 |
| 7 | 2487.20 | −5.57 | 2488.86 | 2431.92 |
| 8 | 2795.75 | −5.65 | 2798.49 | 2725.36 |
| 9 | 3098.66 | −5.75 | 3102.90 | 3011.55 |
| 10 | 3395.81 | −5.85 | 3402.08 | 3290.48 |
| 11 | 3687.12 | −5.96 | 3696.03 | 3562.17 |
| 12 | 3972.47 | −6.09 | 3984.75 | 3826.60 |
| 13 | 4251.72 | −6.22 | 4268.25 | 4083.78 |
| 14 | 4524.76 | −6.38 | 4546.52 | 4333.71 |
| 15 | 4791.41 | −6.54 | 4819.57 | 4576.39 |
| 16 | 5051.53 | −6.72 | 5087.38 | 4811.82 |
| 17 | 5304.92 | −6.92 | 5349.97 | 5039.99 |
| 18 | 5551.38 | −7.14 | 5607.34 | 5260.92 |
| 19 | 5790.71 | −7.38 | 5859.47 | 5474.59 |
| 20 | 6022.64 | −7.65 | 6106.38 | 5681.01 |
| 21 | 6246.93 | −7.94 | 6348.06 | 5880.18 |
| 22 | 6463.28 | −8.25 | 6584.52 | 6072.10 |
| 23 | 6671.38 | −8.60 | 6815.75 | 6256.77 |
| 24 | 6870.88 | −8.98 | 7041.75 | 6434.18 |
| 25 | 7061.40 | −9.39 | 7262.52 | 6604.35 |
| 26 | 7242.53 | −9.83 | 7478.07 | 6767.26 |
| 27 | 7413.84 | −10.32 | 7688.39 | 6922.93 |
| 28 | 7574.82 | −10.85 | 7893.48 | 7071.34 |
| 29 | 7724.96 | −11.41 | 8093.35 | 7212.50 |
| 30 | 7863.69 | −12.02 | 8287.99 | 7346.40 |
| 31 | 7990.39 | −12.65 | 8477.40 | 7473.06 |
| 32 | 8104.45 | −13.31 | 8661.58 | 7592.47 |
| 33 | 8205.20 | −13.95 | 8840.54 | 7704.62 |
| 34 | 8292.00 | −14.53 | 9014.27 | 7809.52 |
| 35 | 8364.28 | −14.94 | 9182.78 | 7907.17 |
| 36 | 8421.61 | −14.98 | 9346.05 | 7997.57 |
| 37 | 8463.95 | −14.27 | 9504.10 | 8080.72 |
| 38 | 8492.03 | −12.28 | 9656.93 | 8156.62 |
| 39 | 8507.83 | −8.87 | 9804.52 | 8225.26 |
| 40 | 8514.75 | −5.05 | 9946.89 | 8286.66 |
| 41 | 8516.62 | | 10084.03 | 8340.80 |
| 42 | | | 10215.95 | 8387.69 |
| 43 | | | 10342.64 | 8427.33 |
| 44 | | | 10464.10 | 8459.72 |
| 45 | | | 10580.33 | 8484.86 |
| 46 | | | 10691.34 | 8502.75 |
| 47 | | | 10797.12 | 8513.38 |
| 48 | | | 10897.67 | 8516.76 |
| 49 | | | 10993.00 | |
| 50 | | | 11083.10 | |
| 51 | | | 11167.97 | |
| 52 | | | 11247.61 | |
| 53 | | | 11322.03 | |
| 54 | | | 11391.22 | |
| 55 | | | 11455.19 | |
| 56 | | | 11513.92 | |
| 57 | | | 11567.43 | |
| 58 | | | 11615.72 | |
| 59 | | | 11658.77 | |
| 60 | | | 11696.60 | |
| 61 | | | 11729.20 | |
| 62 | | | 11756.58 | |
| 63 | | | 11778.73 | |
| 64 | | | 11795.65 | |
| 65 | | | 11807.34 | |
| 66 | | | 11813.81 | |
| 67 | | | 11815.05 | |

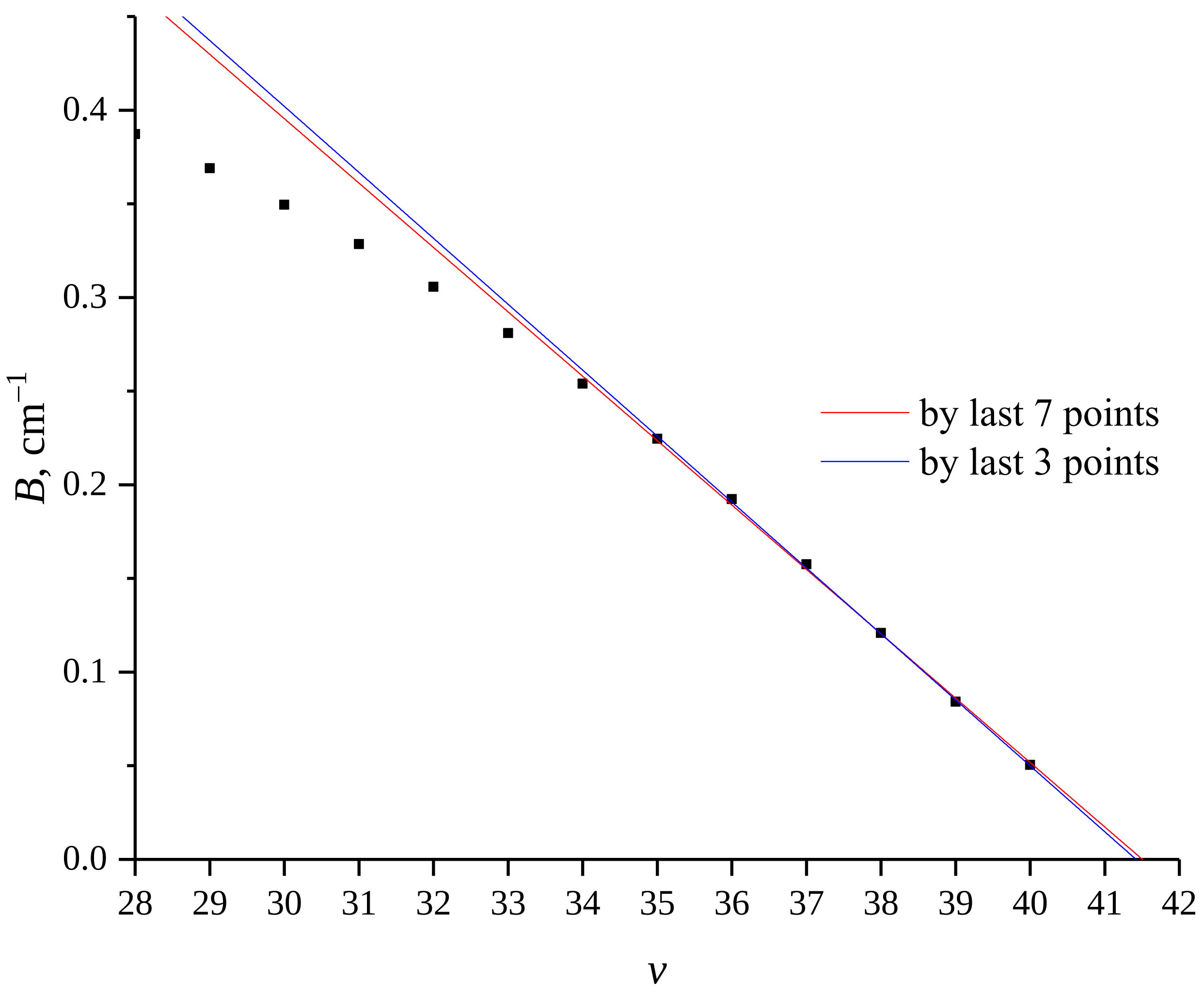

**Figure S2**. The rotational constant $B$ vs vibrational number $v$ for the term X $^1\Sigma_g^+$ of the $Li_2$ molecule according to [4]. The red line is linear fit by last seven points ($v$ = 34…40), the blue line is linear fit by last three points ($v$ = 37…40).

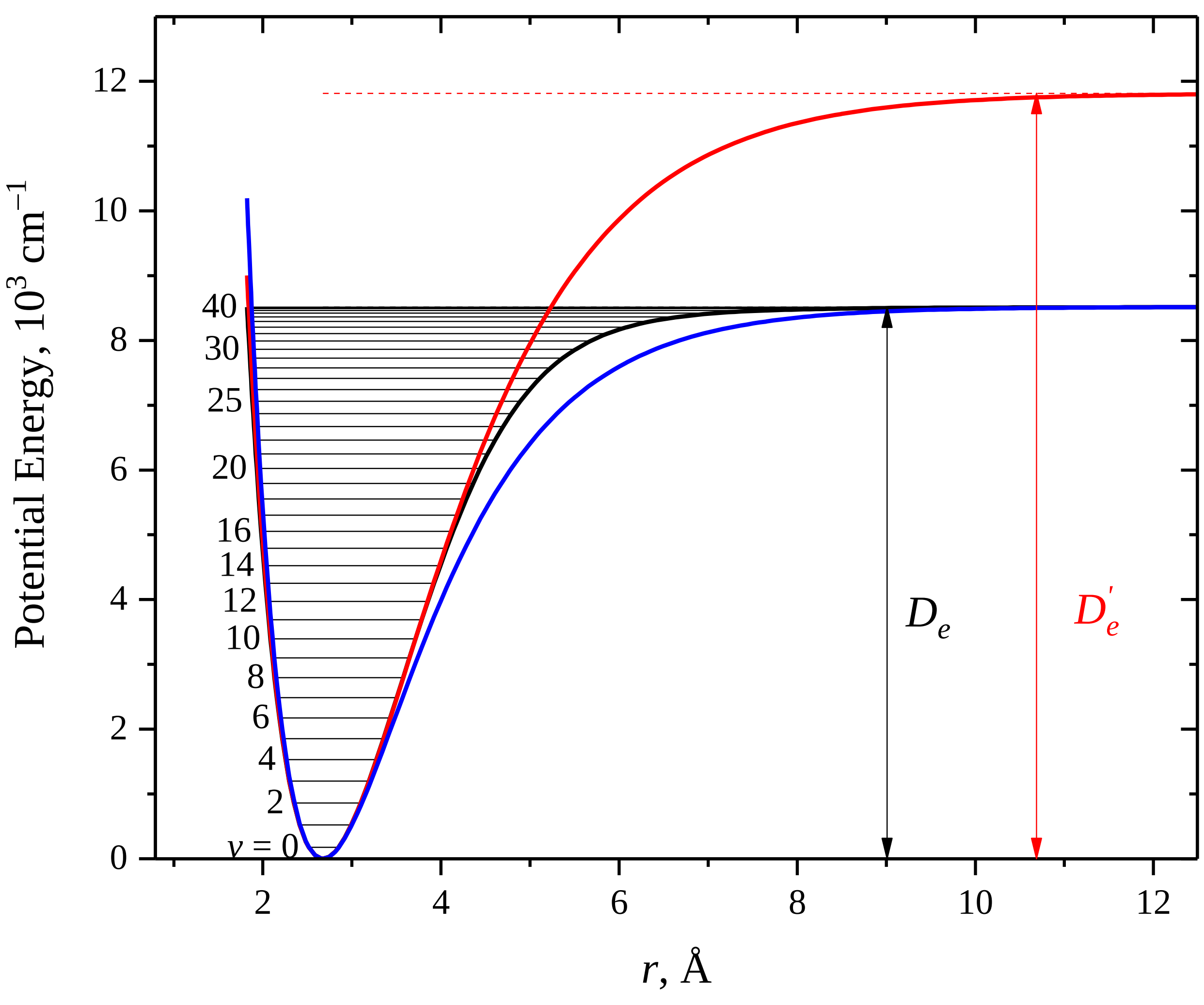

**Figure S3**. The term X $^1\Sigma_g^+$ of the $Li_2$ molecule (black line) according to [4] and its Morse approximations $M1(r)$ (red) and $M2(r)$ (blue). $D_e$ = 8517 cm$^{-1}$, $D_e'$ = 11815 cm$^{-1}$. There are 27 fictitious levels for *M*1 and 8 for *M*2.

**$B_2$**. The potential $U(r)$ for the term X $^3\Sigma_g^-$ of the $^{11}B_2$ molecule were built according the equations in the Table 2 of the article [5] with fitted parameters from this table. It showed total agreement with the author's potential in the points tabulated in Table 3 of the article. The vibrational energies, the values of $D_e$ and $r_e$ were taken from Table 5 of the article. The parameters of M1 and M2 potentials are presented in **Table S7** and energies of vibrational levels in **Table S8**.

**Table S7**. The parameters of the Morse potentials M1 and M2 for X $^3\Sigma_g^-$ term of the $^{11}B_2$ molecule

| Parameter | M1 | M2 |
|---|---|---|
| $\mu$, Da | 5.504652584 | |
| $r_e$, Å | 1.5886 | |
| $D_e$,[a] cm$^{-1}$ | 31482.6 | 23686.3 |
| $\omega_e$ , cm$^{-1}$ | 1049.41 | |
| $\omega_e x_e$,[b] cm$^{-1}$ | 8.745 | 11.623 |
| $x_e$[c] | 0.00833 | 0.01108 |
| $a$, Å$^{-1}$ | 1.6898 | 1.9482 |

[a] $D_e'$ for M1; [b] $\omega_e x_e'$ for M2; [c] $x_e'$ for M2.

**Table S8**. The energies (cm$^{-1}$) of levels $v$ for potentials $U(r)$ of the X $^1\Sigma_g^+$ state of the $^{11}B_2$ molecule [5], and its approximations $M1$ and $M2$. The anharmonicity function $-2\omega_e x$ (cm$^{-1}$) of $U(r)$ is presented in the third column.

| $v$ | $U$ | $-2\omega_e x$ | M1 | M2 |
|---|---|---|---|---|
| 0 | 522.52 | | 522.52 | 521.80 |
| 1 | 1554.44 | −17.49 | 1554.44 | 1547.96 |
| 2 | 2568.87 | −17.71 | 2568.87 | 2550.88 |
| 3 | 3565.59 | −17.92 | 3565.81 | 3530.55 |
| 4 | 4544.39 | −18.16 | 4545.26 | 4486.97 |
| 5 | 5505.03 | −18.42 | 5507.22 | 5420.15 |
| 6 | 6447.25 | −18.70 | 6451.69 | 6330.08 |
| 7 | 7370.77 | −18.99 | 7378.67 | 7216.76 |
| 8 | 8275.30 | −19.31 | 8288.16 | 8080.19 |
| 9 | 9160.52 | −19.65 | 9180.16 | 8920.38 |
| 10 | 10026.09 | −20.01 | 10054.67 | 9737.33 |
| 11 | 10871.65 | −20.42 | 10911.69 | 10531.02 |
| 12 | 11696.79 | −20.82 | 11751.22 | 11301.47 |
| 13 | 12501.11 | −21.28 | 12573.26 | 12048.67 |
| 14 | 13284.15 | −21.77 | 13377.81 | 12772.63 |
| 15 | 14045.42 | −22.29 | 14164.87 | 13473.33 |
| 16 | 14784.40 | −22.85 | 14934.44 | 14150.79 |
| 17 | 15500.53 | −23.48 | 15686.52 | 14805.01 |
| 18 | 16193.18 | −24.13 | 16421.11 | 15435.98 |
| 19 | 16861.70 | −24.85 | 17138.21 | 16043.70 |
| 20 | 17505.37 | −25.65 | 17837.82 | 16628.17 |
| 21 | 18123.39 | −26.49 | 18519.94 | 17189.40 |
| 22 | 18714.92 | −27.44 | 19184.57 | 17727.38 |
| 23 | 19279.01 | −28.48 | 19831.71 | 18242.11 |
| 24 | 19814.62 | −29.61 | 20461.36 | 18733.60 |
| 25 | 20320.62 | −30.89 | 21073.52 | 19201.84 |
| 26 | 20795.73 | −32.26 | 21668.19 | 19646.83 |
| 27 | 21238.58 | −33.80 | 22245.37 | 20068.58 |
| 28 | 21647.63 | −35.46 | 22805.06 | 20467.08 |
| 29 | 22021.22 | −37.22 | 23347.26 | 20842.33 |
| 30 | 22357.59 | −38.98 | 23871.97 | 21194.34 |
| 31 | 22654.98 | −40.48 | 24379.19 | 21523.10 |
| 32 | 22911.89 | −41.25 | 24868.92 | 21828.61 |
| 33 | 23127.55 | −40.43 | 25341.16 | 22110.87 |
| 34 | 23302.78 | −37.24 | 25795.91 | 22369.89 |
| 35 | 23440.77 | −32.15 | 26233.17 | 22605.67 |
| 36 | 23546.61 | −27.67 | 26652.94 | 22818.19 |
| 37 | 23624.78 | −26.92 | 27055.22 | 23007.47 |
| 38 | 23676.03 | | 27440.01 | 23173.50 |
| 39 | | | 27807.31 | 23316.29 |
| 40 | | | 28157.12 | 23435.82 |
| 41 | | | 28489.44 | 23532.11 |
| 42 | | | 28804.27 | 23605.16 |
| 43 | | | 29101.61 | 23654.96 |
| 44 | | | 29381.46 | 23681.51 |
| 45 | | | 29643.82 | 23684.81 |
| 46 | | | 29888.69 | |
| 47 | | | 30116.07 | |
| 48 | | | 30325.96 | |
| 49 | | | 30518.36 | |
| 50 | | | 30693.27 | |
| 51 | | | 30850.69 | |
| 52 | | | 30990.62 | |
| 53 | | | 31113.06 | |
| 54 | | | 31218.01 | |
| 55 | | | 31305.47 | |
| 56 | | | 31375.44 | |
| 57 | | | 31427.92 | |
| 58 | | | 31462.91 | |
| 59 | | | 31480.41 | |
| 60 | | | 31480.42 | |

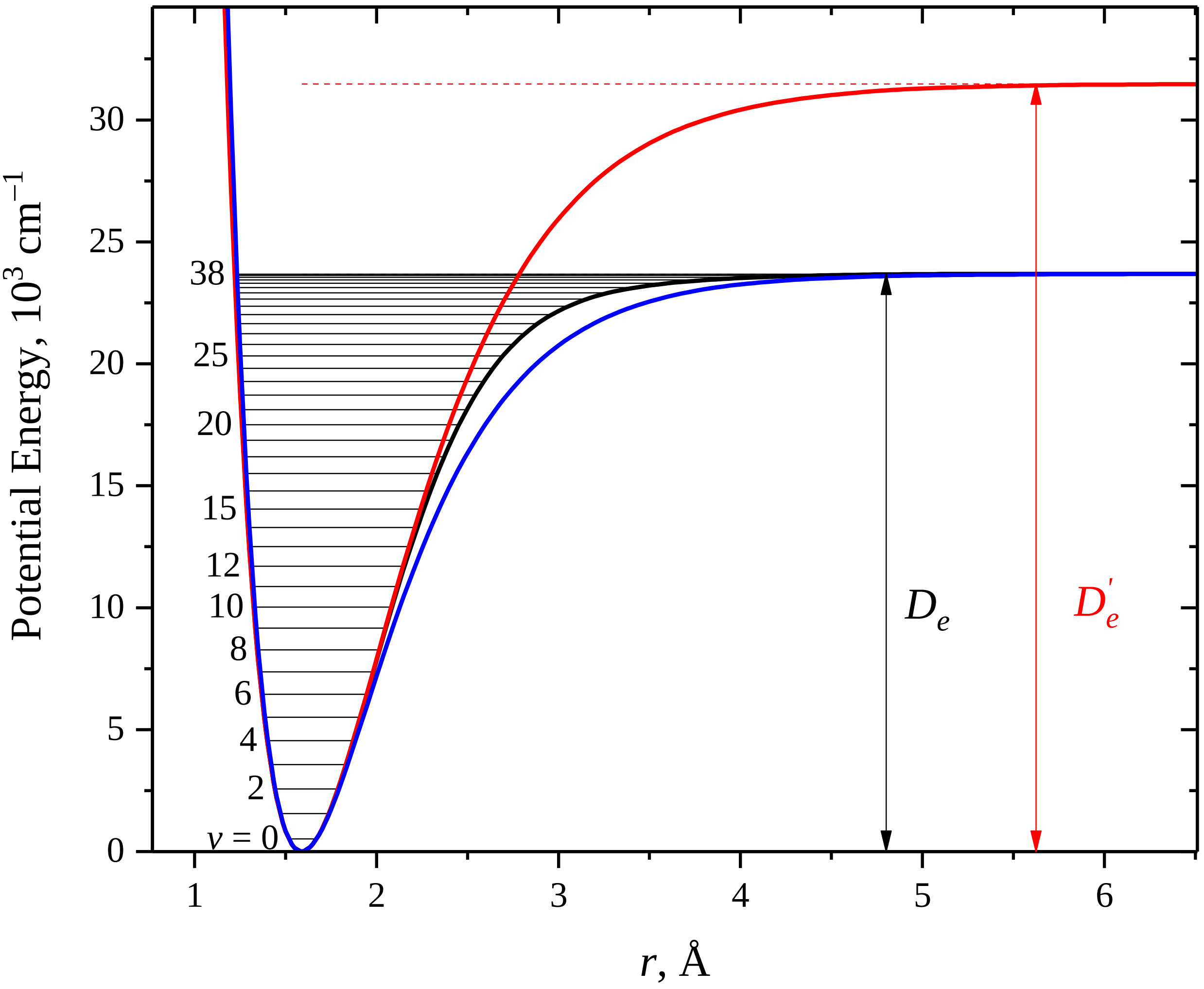

**Figure S4**. The term X $^3\Sigma_g^-$ of the $B_2$ molecule (black line) according to [5] and its Morse approximations $M1(r)$ (red) and $M2(r)$ (blue). $D_e$= 23686.3 cm$^{-1}$, $D_e'$ = 31482.6 cm$^{-1}$. There are 21 fictitious levels for $M1$ and 7 for $M2$.